\documentclass[aps,floatfix,superscriptaddress,amsmath,amssymb,noshowpacs,twocolumn]{revtex4-1}

\usepackage[USenglish]{babel}
\usepackage{xcolor}
\usepackage{array}

\usepackage{amsmath,amssymb,amsfonts,mathrsfs}

\usepackage{bm}

\usepackage{epsfig}

\usepackage[colorlinks=true,linkcolor=blue,citecolor=red]{hyperref}

\newcommand{\be}{\begin{equation}}
\newcommand{\ee}{\end{equation}}
\newcommand{\beS}{\begin{equation*}}
\newcommand{\eeS}{\end{equation*}}
\newcommand{\bea}{\begin{eqnarray}}
\newcommand{\eea}{\end{eqnarray}}
\newcommand{\ba}{\begin{eqnarray*}}
\newcommand{\ea}{\end{eqnarray*}}
\newenvironment{eqs}
{\begin{equation} \begin{aligned}}
{\end{aligned} \end{equation} }
\newcommand{\bal}{\begin{eqs}}
\newcommand{\eal}{\end{eqs}}

\newcommand{\bas}{\begin{eqs}}
\newcommand{\eas}{\end{eqs}}

\newcommand{\Tr}{\text{Tr}}

\newcommand{\bw}{\begin{widetext}}
\newcommand{\ew}{\end{widetext}}

\begin{document}
\title{Probing Majorana edge states by measuring transport through an interacting magnetic impurity}
\author{Daniele Guerci}
\affiliation{International School for
  Advanced Studies (SISSA), Via Bonomea
  265, I-34136 Trieste, Italy} 
\author{Andrea Nava}
\affiliation{International School for
  Advanced Studies (SISSA), Via Bonomea
  265, I-34136 Trieste, Italy} 

\date{\today} 
\pacs{}
\begin{abstract}

Motivated by recent experiments we consider transport across an interacting magnetic impurity coupled to the Majorana zero mode (MZM) observed at the boundary of a topological superconductor (SC). In the presence of a finite tunneling amplitude we observe hybridization of the MZM with the quantum dot, which is manifested by a half-integer zero-bias conductance $G_0=e^2/2h$ measured on the metallic contacts.
The low-energy feature in the conductance drops abruptly by crossing the transition line from the topological to the non-topological superconducting regime.
Differently from the in-gap Yu-Shiba-Rosinov-like bound states, which are strongly affected by the on-site impurity Coulomb repulsion, we show that the MZM signature in the conductance is robust and persists even at large values of the interaction. Interestingly, the topological regime is characterized by a vanishing Fano factor, $F=0$, induced by the MZM. Combined measurements of the conductance and the shot noise in the experimental set-up presented in Fig. \ref{system3d} allow to detect the topological properties of the superconducting wire and to distinguish the low-energy contribution of a MZM from other possible sources of zero-bias anomaly.
Despite being interacting the model is exactly solvable, which allows to have an exact characterization of the charge transport properties of the junction. 

\end{abstract}
\maketitle

\paragraph*{Introduction.}

After the seminal paper by Kitaev Ref. \cite{kitaev_unpaired_2001} that predicted the existence of electronic collective modes reminiscent of the Majorana fermions speculated in 1937 by Ettore Majorana \cite{Majorana2008}, quasi one-dimensional systems, hosting two or more Majorana zero modes (MZMs), have attracted both experimental \cite{Oreg_2012,Mourik1003,Albrecht_2016,Deng_2016,Kouwenhoven_MZM_2018,Kouwenhoven_Nature_2018_QC,Kouwenhoven_Nature_2018_BM,Flensberg_2017,He294,EstradaSaldanaeaav1235} and theoretical \cite{PhysRevLett.105.177002,PhysRevLett.100.096407,PhysRevLett.86.268,PhysRevB.61.10267} interest. 
Intrigued by exciting prospects in fault-tolerant quantum computation \cite{RevModPhys.80.1083,Alicea2011,Alicea_2012}, existing theoretical studies focused on zero-bias and current measurement across a junction of metallic leads and topological superconductors (SCs) \cite{Affleck_2013,GIULIANO2019114645,Affleck2014,PhysRevB.85.245121,Zhang_Maciejko_2011,Pientka_2015,Klinovaja_2018},  shot noise measurement \cite{Chevallier_2017,Giuliano_shotnoise}, interferometer measurement \cite{PhysRevB.94.125426}, persistent current in hybrid normal-superconducting rings \cite{PhysRevB.88.241409,Pientka_2013,PhysRevB.94.205125,PhysRevB.95.155449} and topological realization of the Kondo effect \cite{PhysRevLett.109.156803,PhysRevLett.113.076404,PhysRevB.90.245417}.

Recently, a new direction has emerged which explores the interplay between pure Majorana physics and electronic correlations \cite{Ng_2018_Majorana,PhysRevLett.115.166401,PhysRevB.92.235123,PhysRevB.96.125121}. 
 
In this letter we fully characterize the electronic transport through a novel class of experimentally realizable systems \cite{Deng_2016,Kouwenhoven_Nature_2018_QC} which have recently attracted great interest for their easily realization and control. The MZM, emerging at the endpoint of a one dimensional semi-infinite wire with strong spin-orbit interaction (i.e. InAs wire) deposited on top of a s-wave SC and exposed to an external magnetic field, is coupled to an interacting magnetic impurity that can be used as a spectrometer. By coupling the dot to two metallic fully-polarized contacts we can probe the properties of the MZM though measurement of the current and the shot noise across the junction.

\begin{figure}
\begin{center}
\includegraphics[width=0.4\textwidth]{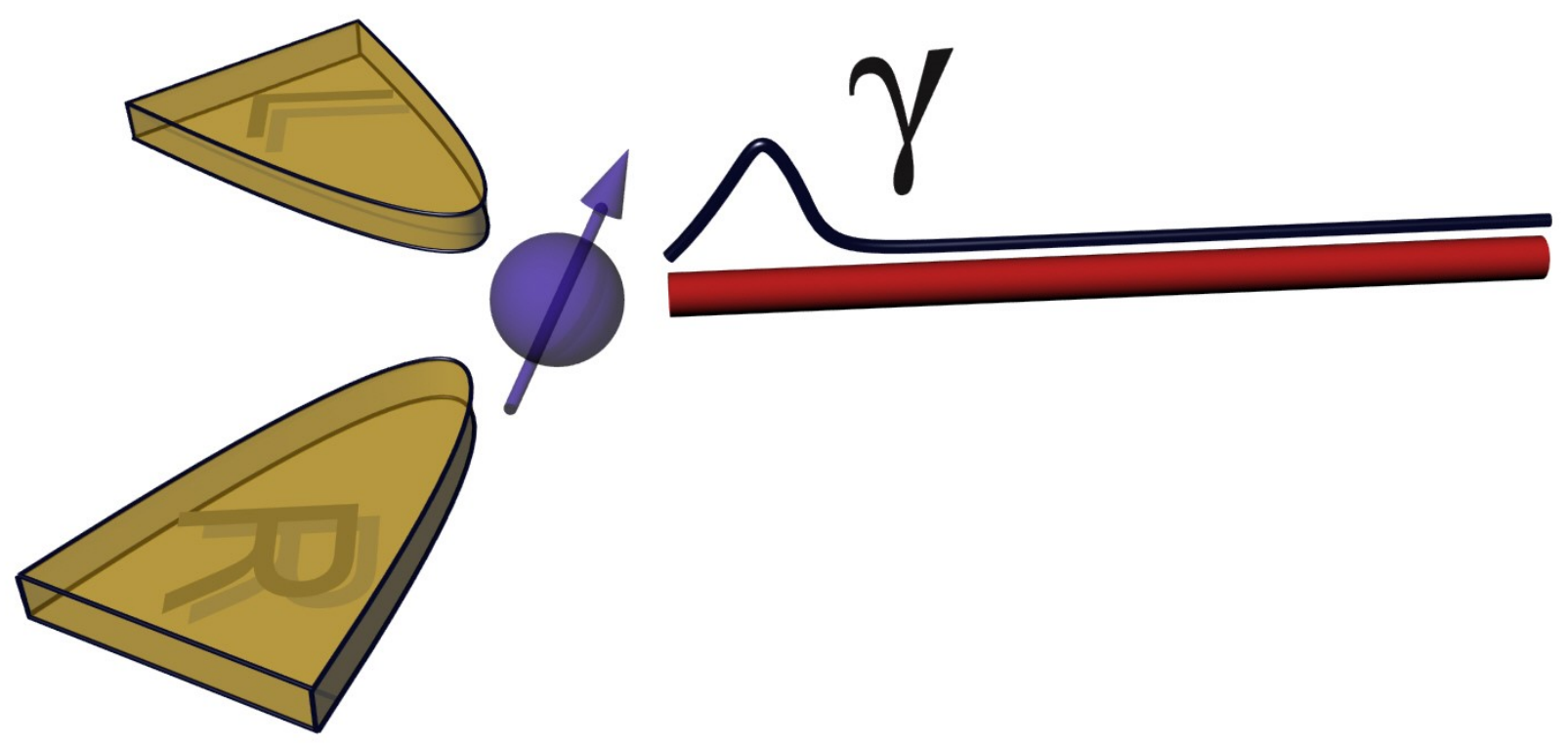}
\vspace{-0.05cm}
\caption{Sketch of a quantum dot coupled to two fully-polarized metallic leads and a semi-infinite topological p-wave SC hosting a MZM at its edge.}
\label{system3d}
\end{center}
\end{figure}

\paragraph*{Model Hamiltonian.}
To model the junction displayed in Fig. \ref{system3d} we consider the Hamiltonian
\be
\label{model_junction}
H=H_{\text{imp}}+H_{C}+H_{K}+H_{T,C}+H_{T,K},
\ee
where 
\be
\label{atomic_H}
H_{\text{imp}}=\frac{U}{4}\Omega_d-\frac{h}{2}\left(n^d_\uparrow-n^d_\downarrow\right)+\frac{\mu}{2}\left(n^d_\uparrow+n^d_\downarrow-1\right)
\ee
is the dot Hamiltonian, with $n^d_\sigma=d^\dagger_\sigma d_\sigma$ the number operator on the impurity site and $\Omega_d=(2n^d_\uparrow-1)(2n^d_\downarrow-1)$. In (\ref{atomic_H}) $U$ denotes the on-site interaction, $\mu$ the gate potential and $h$ the Zeeman field applied on the dot level.  The Hamiltonian of the semi-infinite Kitaev chain reads
\be
\label{H_chain}
H_{K}=\sum_{j=1}^{\infty}[(-t c^\dagger_{j}c_{j+1}+\Delta c_{j}c_{j+1}+\text{H.c.})-\mu c^\dagger_{j}c_{j}]
\ee 
 where $t$ is the hopping amplitude between nearest neighbor sites, $\Delta$ the p-wave superconducting pairing and $\mu$ the chemical potential of the wire.
 We notice that left $(L)$ and right $(R)$ metallic contacts are described by Hamiltonian (\ref{H_chain}) with $\Delta=0$ and different electrochemical potentials $\mu_{L}=-\mu_R=\phi/2$. In our model both the Kitaev and the metallic chains are described by spinless particles. This is a natural assumption if one consider that topological SCs are realized in one dimensional p-wave SCs characterized by strong spin-orbit coupling and large magnetic fields, and if we assume fully-polarized ferromagnetic contacts. In this regime the magnetic exchange between the impurity spin and the leads is suppressed and the low-energy physics is dominated by the coupling with the MZM \cite{Nagaosa_MZM_2010,Pascal_MZM_2011,Rosa_2013,Cheng_2014,Weymann_2017}.
 The tunneling between the dot and the metallic contacts reads:
 \be
\label{H_T_mcontacts}
H_{T,C}=V_{c}\sum_{\alpha=L,R}\left(c^\dagger_{1\alpha}d_\uparrow+\text{H.c.}\right)
\ee 
where $V_c$ is the tunneling amplitude and $\alpha=L,R$. Finally, we consider the hybridization with the boundary site of the semi-infinite Kitaev chain: 
 \be
\label{H_T_Kitaev}
H_{T,K}=-i\sum_{j}V_j\gamma_{j}\gamma^{d}_\uparrow, 
\ee
where the sum extends to the semi-infinite Kitaev chain and we have introduced the Majorana operators $\gamma=c+c^\dagger$ and $\xi=-i(c-c^\dagger)$.
The simple model in Eq. (\ref{H_T_Kitaev}) allows to study exactly the effect of correlations on the non-local Majorana edge state tunnel-coupled to an interacting quantum dot.
   
The interacting model is exactly solvable because the $d_{\downarrow}$ electrons are localized and $n^d_{\downarrow}$ can be treated as a $\mathbb{Z}_{2}$ real number $(=0,1)$. This property makes the Hamiltonian (\ref{model_junction}) an effective quadratic model, where similarly to the Falicov-Kimball model (FKM) \cite{Falicov_1969} the $\downarrow$ configuration is obtained by minimizing the ground-state energy of the $\uparrow$ degrees of freedom.

In the absence of metallic contacts, $V_c=0$, the equilibrium properties of the model in Eq. (\ref{model_junction}) has been already studied in Ref. \cite{Shankar_MAI}. It is convenient to perform the following gauge transformation:
\be
\label{GAUGE_xi}
\xi^\eta_\uparrow=\xi^d_\uparrow (1-2n^d_{\downarrow}),\quad\gamma^\eta_\uparrow=\gamma^d_\uparrow,
\ee
in terms of $\gamma^\eta_{\uparrow}$ and $\xi^\eta_{\uparrow}$ fermions the Hamiltonian (\ref{model_junction}) becomes \cite{comment_Shankar}:  
\bal
\label{model_junction_SS_A}
H^{*}=&H_{C}+H_{K}-i\sum_{j}V_j\gamma_{j}\gamma^{\eta}_\uparrow\\
&-i\frac{U}{4}\gamma^\eta_\uparrow\xi^\eta_\uparrow-\frac{(\mu+h)-i(\mu-h)\gamma^\eta_\uparrow\xi^\eta_\uparrow}{4}q^d_{\downarrow}\\
&+i\frac{V_c}{2}\sum_{\alpha=L,R}\left(\gamma^\eta_\uparrow\xi_{1\alpha}-q^d_{\downarrow}\xi^\eta_\uparrow\gamma_{1\alpha}\right).
\eal
To avoid irrelevant complications we consider the case $\mu=h=0$. Introducing the Dirac (complex) fermion $\eta_{\uparrow}=\gamma^\eta_\uparrow+i\xi^\eta_\uparrow$, the model Hamiltonian reads 
\bal
\label{model_junction_SS_B}
H^{*}&=H_{C}+H_{K}+\frac{1}{2}\sum_{\alpha=L,R}\left(\vec{\eta}^\dagger_\uparrow\cdot\hat{V}_c\cdot\vec{c}_{1\alpha}+\text{H.c.}\right)\\
&+\frac{1}{2}\sum_j\left(\vec{\eta}^\dagger_\uparrow\cdot\hat{V}_j\cdot\vec{c}_{j}+\text{H.c.}\right)-\frac{1}{2}\vec{\eta}^\dagger_\uparrow\cdot\frac{U}{2}\sigma^z\cdot\vec{\eta}_\uparrow,
\eal
where in the Nambu representation $\vec{\psi}=(\psi,\,\,\psi^\dagger)^T$, $\hat{V}_{j}$ is the hybridization matrix between the dot and the $j$-th site of the Kitaev chain:
\be
  \label{V_hopping}
  \hat{V}_j=iV_j\left(\begin{array}{cc}1 & 1 \\1 & 1\end{array}\right),
\ee
and $\hat{V}_c$ couples the metallic contacts to the dot
\be
  \label{V_metal}
  \hat{V}_{c}=\frac{V_c}{2}\left(\begin{array}{cc}(1+q^d_\downarrow) & -(1-q^d_\downarrow) \\(1-q^d_\downarrow) & -(1+q^d_\downarrow)\end{array}\right).
\ee

To characterize the transport properties of the junction we compute the charge current, $J_Q=(J_L-J_R)/2$ with $J_{\alpha}=-i[N_\alpha,H]$, that in the new representation (\ref{GAUGE_xi}) reads:
\be
\label{current}
J_{Q}=-i\frac{V_c}{4}\sum_{\alpha=L,R}\text{sign}(\alpha)\left[\gamma_{1\alpha}\gamma^\eta_\uparrow+q^d_\downarrow\xi_{1\alpha}\xi^{\eta}_\uparrow\right]
\ee
where $\text{sign}(L)=+1$ and $\text{sign}(R)=-1$, and the zero frequency limit of the $J_Q$ fluctuations 
\be
\label{shot_noise}
S_{Q}=\int d(t-t^\prime) \frac{\langle\{\delta J_Q(t),\delta J_Q(t^\prime)\}\rangle}{2},
\ee
where $\delta J_Q= J_Q-\langle J_Q\rangle$. 
In the following we study transport through the junction by performing calculations with Keldysh Green's function technique \cite{rammer_2007,HAUG_JAUHO}, which we compare with the scattering matrix approach \cite{BLANTER_2001,Nazarov_Blanter,Nilsson}.

\paragraph*{Probing MZMs with charge conductance and shot noise.}
Experimental measurements of charge conductance at the boundary of topological materials reveal the emergence of low-energy MZMs \cite{Oreg_2012,Albrecht_2016,Deng_2016,Kouwenhoven_MZM_2018,Kouwenhoven_Nature_2018_QC,Kouwenhoven_Nature_2018_BM,Flensberg_2017} and provide an experimental tool to detect topological transitions by studying surface states via STM \cite{Bernevig_MZM_2019,Bernevig_2018_STM,STM_Schindler_2018,STM_Drozdov_2014,Josephson_Murani_2017}.
  \begin{figure}
\begin{center}
\includegraphics[width=0.48\textwidth]{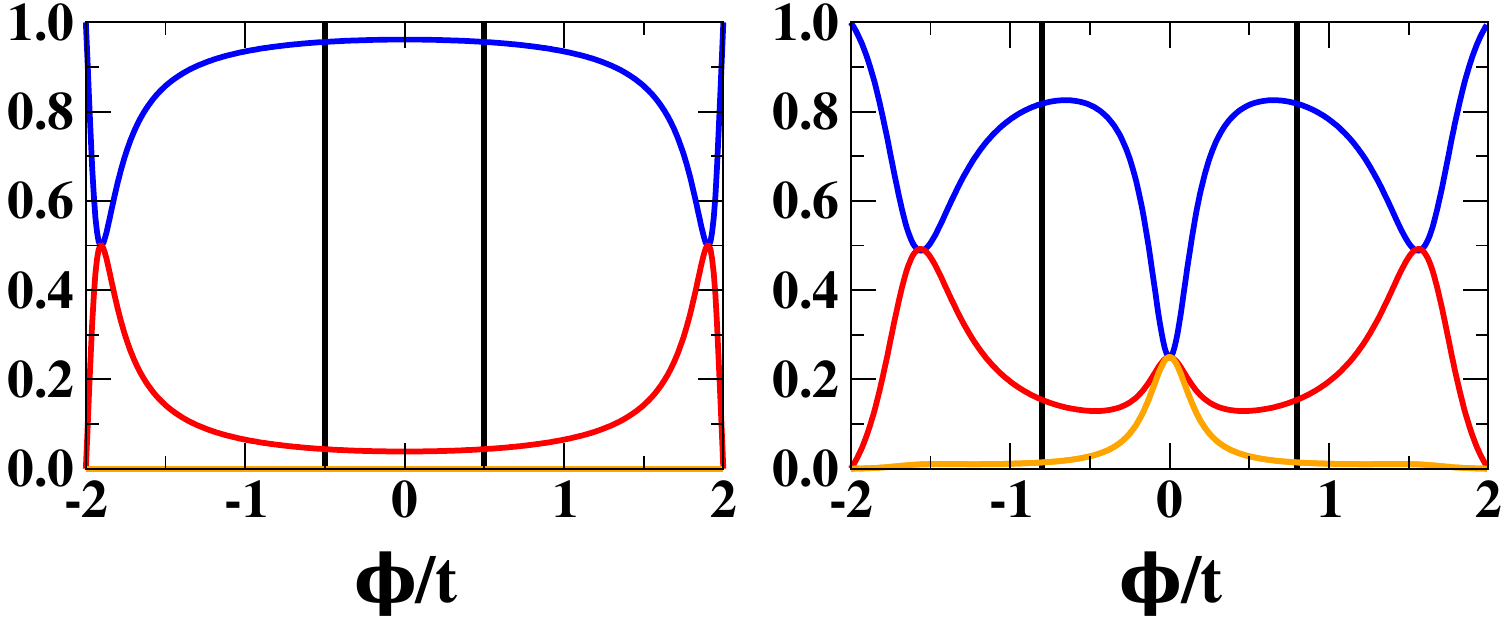}
\vspace{-0.5cm}
\caption{ Scattering matrix coefficients for the system in Fig. \ref{system3d} in the trivial regime (left panel) and topological regime (right panel) at finite interaction $U/t=1.6$. Blue: normal reflection; red: normal transmission, orange: Andreev reflection and crossed Andreev reflection. Vertical black lines show the "bulk" superconducting gap $\Delta_\text{gap}$.}
\label{S-matrix}
\end{center}
\end{figure}

 In this letter we present a detailed characterization of the low-energy signatures observed in the charge conductance and shot noise measurements that allows to classify different regions of the Kitaev chain phase diagram. Despite being done on a toy model, the analysis may give physical insight for the understanding of the ongoing experiments where the effect of local on-site interaction cannot be neglected.
We start by reporting the expression of the current flowing through the metallic contacts:
\bal
\label{current}
\left\langle J_Q\right\rangle=&\frac{\pi e^2}{h}\int d\epsilon\bar{\rho}(\epsilon)\left(f_L(\epsilon)-f_R(\epsilon)\right)\text{Im}\Tr\left(\hat{\mathbf{T}}^A_{\eta_\uparrow}(\epsilon)\right)
\eal
 where $f_L(\epsilon)=f(\epsilon-\phi/2)$, $f_R(\epsilon)=f(\epsilon+\phi/2)$, $\Tr$ is the trace in the $2\times2$ Nambu space, $\hat{\mathbf{T}}^{R/A}(\epsilon)$ is the impurity transfer matrix
 \be
 \label{impurity_transfer}
 \hat{\mathbf{T}}^{R/A}(\epsilon)=\hat{V}^\dagger_{c}\cdot \hat{\mathbf{G}}^{R/A}_{\eta_\uparrow}(\epsilon)\cdot \hat{V}_{c},
 \ee
  and $\bar{\rho}(\epsilon)$ is the boundary density of states for the semi-infinite normal contacts (we refer to the supplemental material for more details \cite{supplementary}). 
  \begin{figure}
\begin{center}
\includegraphics[width=0.48\textwidth]{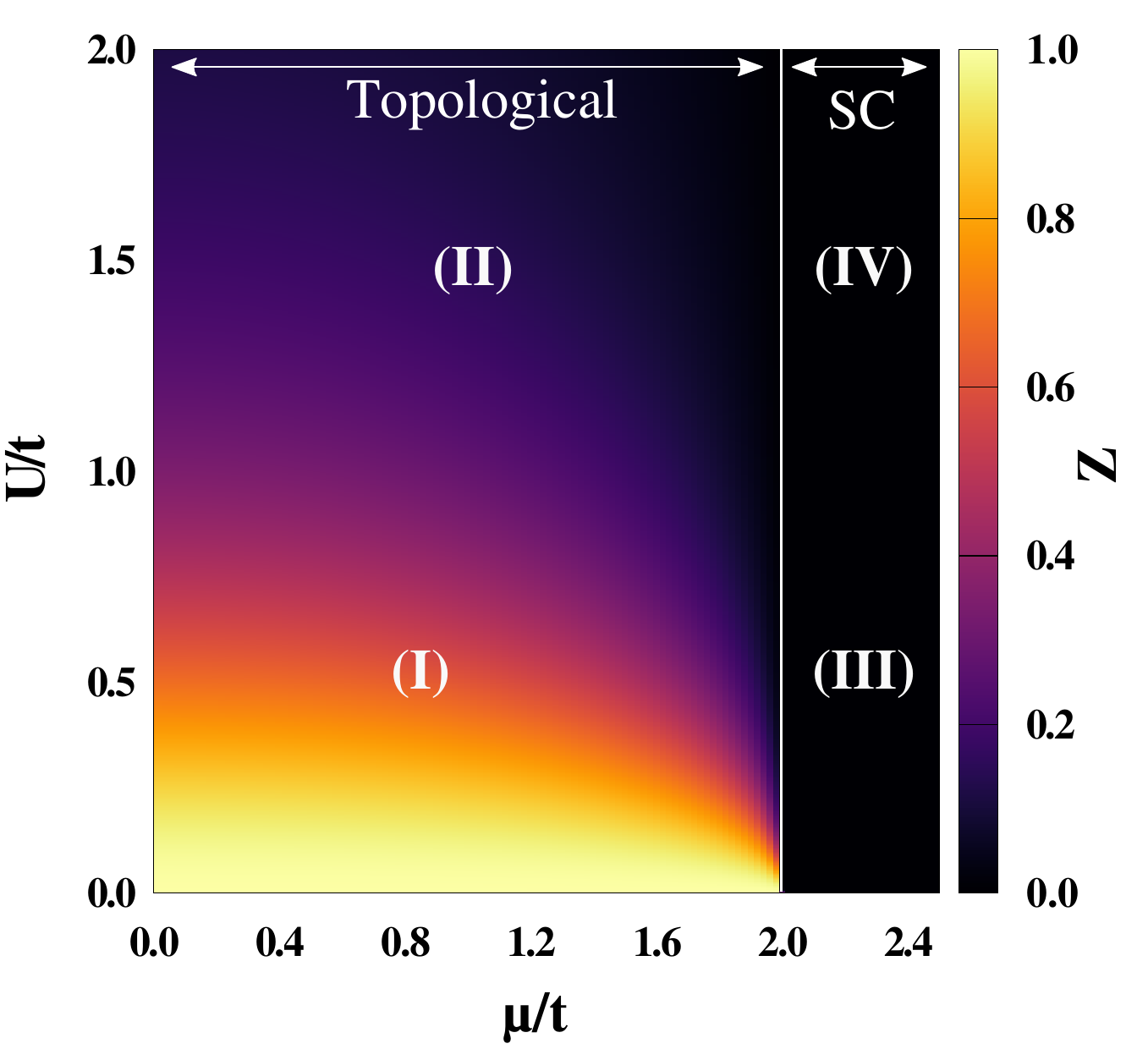}
\vspace{-0.5cm}
\caption{ Quasiparticle renormalization factor $Z$ of the low-energy MZM as a function of $\mu/t$ and $U/t$, for $\Delta/t=V/t=0.4$. Symbols from (I) to (IV) characterize different charge transport behavior, see Fig. (\ref{JG_phases}).}
\label{zeta_Uvsmu}
\end{center}
\end{figure}
The resulting value of the current is obtained by averaging over the spin $\downarrow$ configurations:
\be
\label{charge_spin_av}
\langle\langle J_Q\rangle\rangle=\sum_{n^f_\downarrow=0,1}p(n^f_\downarrow)\langle J_Q(n^f_\downarrow)\rangle
\ee
where in the absence of any gate potential or Zeeman field on the quantum dot $p(0)=p(1)=1/2$.
 
In the topological regime, the low-energy physics is governed by the in-gap states that emerge from the hybridization between the real and imaginary part of the spin up dot fermion and the MZM of the Kitaev chain. The coupling between $\gamma^d_{\uparrow}$ and $\gamma_1$ induces an energy splitting $\sim V$, while the quantum dot interaction generates an energy splitting $\sim U$ between $\gamma^d_{\uparrow}$ and $\xi^d_{\uparrow}$. The combined effect of the dot-Kitaev chain coupling and the interaction, on an odd number of MZMs, is to split two of them by a term $\sim f(U,V)$ that eventually, for $U$ strong enough, wash them out from the superconducting gap. Whereas, the third one is a topologically protected, and robust to the interaction, zero energy mode. In the trivial regime, we have an even number of MZMs, then no zero energy mode is preserved as any finite interaction induces a hybridization $\sim U$ between them.

 \begin{figure}
\begin{center}
\includegraphics[width=0.48\textwidth]{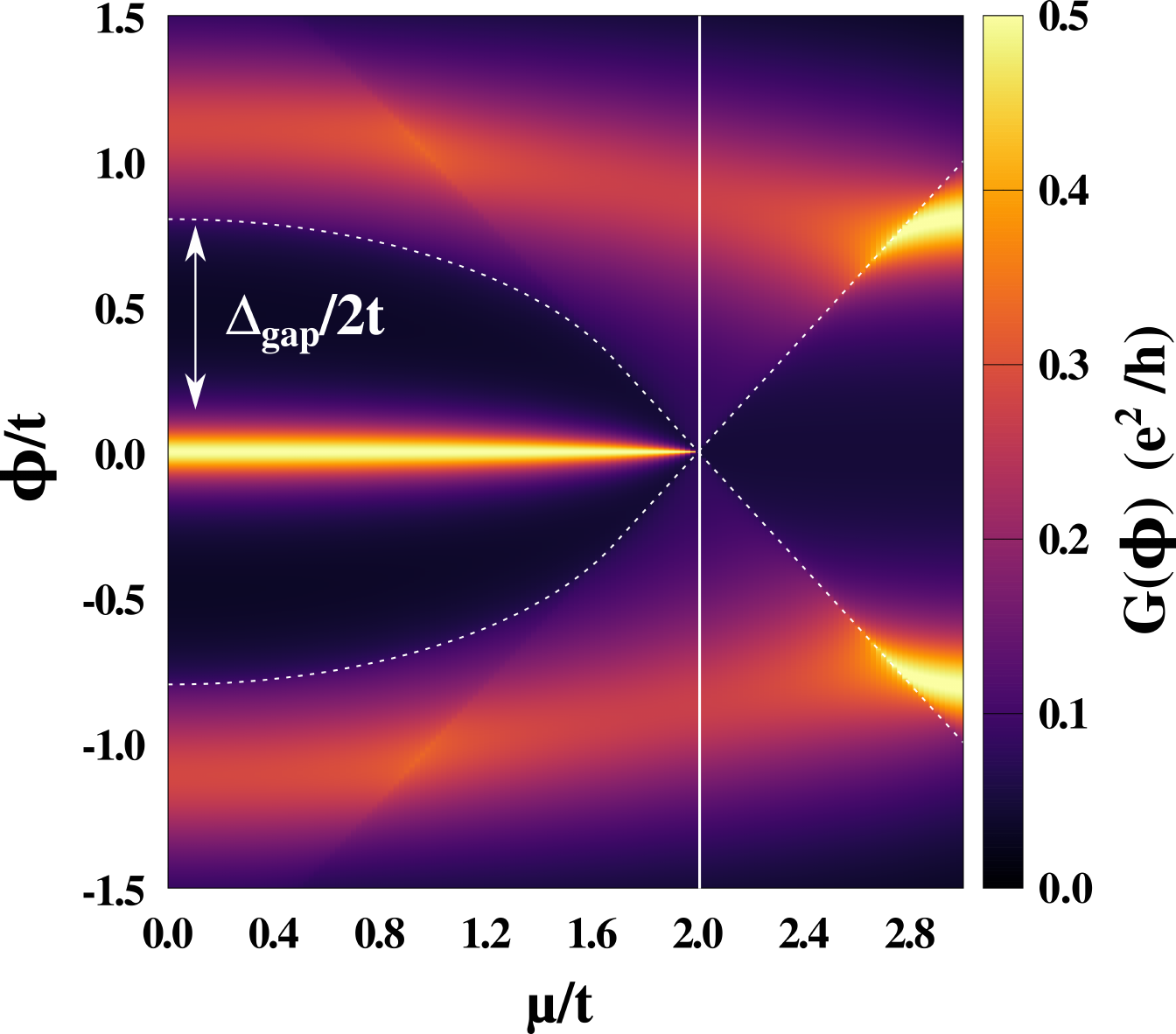}
\vspace{-0.5cm}
\caption{ Evolution of the  conductance $G(\phi)$ as a function of $\mu/t$, for $U/t=1.6$, $\Delta/t=V/t=0.4$ and $V_c/t=0.3$. Dashed white line shows the superconducting gap measured on the boundary site of the semi-infinite Kitaev chain. White vertical line corresponds to the topological transition.}
\label{G_vs_mu}
\end{center}
\end{figure}

These features can by easily detected resorting to the scattering matrix approach of Ref. \cite{Nilsson, Nazarov_Blanter} that allows to interpret the transport properties of the system in terms of the scattering processes across the junction (a detailed description is given in the supplementary material \cite{supplementary}). In the trivial regime (left panel of Fig. \ref{S-matrix}), the presence of massive in-gap modes suppresses low-energy scattering processes so that the $L$ and $R$ contacts are disconnected in the large $U/V_c$ limit. On the contrary, in the topological regime (right panel of Fig. \ref{S-matrix}), the presence of the MZM keeps alive all the scattering processes at low-energy. The normal transmission (T), the Andreev reflection (A) and the crossed Andreev reflection (C) are equal to one fourth at any value of $U$ and $V$. As a consequence, the charge current, $J_Q$, that measures the charge imbalance between left and right lead, is $\propto A+T\sim1/2$ and the zero-bias conductance is reduced from $e^2/h$ to $e^2/2h$, as already observed in previous studies \cite{Dong_2011,Rosa_2013,Rosa_2014,Vernek_2014,Beri_2014}. 
Interestingly, the on-site local repulsion does not modify the result $e^2/2h$ while it affects the curvature of the low-bias conductance by renormalizing the MZM:
\be
\label{low_bias_current}
G(\phi)=\frac{\partial\left\langle J_Q\right\rangle}{\partial\phi}\simeq\frac{e^2}{2h}\left[1-\left(\frac{\phi}{2\Gamma_cZ}\right)^2\right],
\ee
where $\Gamma_c=2\pi\bar{\rho}(0)V^2_c$ is the hybridization with the metallic contacts, $\bar{\rho}(\omega)$ the boundary metallic density of states and $Z$ the quasiparticle renormalization factor. The latter quantity is shown in the color map \ref{zeta_Uvsmu}, where we analyze the evolution of $Z$ in different regions of the phase diagram of the Kitaev chain. 
We stress that Eq. (\ref{low_bias_current}) is valid in the topological regime $|\mu|<2t$ where the SC posses a non-trivial topology and a MZM appears at the edge of the semi-infinite Kitaev chain. Differently, in the region $|\mu|>2t$, the MZM disappears and we enter in the Coulomb blockade regime where the zero-bias conductance is suppressed.

 The topological transition is associated to a drastic variation of the conductance $G(\phi)$. Indeed, as shown in Fig. \ref{G_vs_mu}, by crossing the critical line, $\mu=2t$, we observe a jump from $G_0=e^2/2h$ in the topological region to $G_0\simeq0$ in the trivial one, which allows to distinguish the two different phases. Moreover, we notice that in the non-topological region, for $\mu/t\simeq2.5$, the conductance presents coherent in-gap peaks attributable to Andreev bound states induced by the impurity,  reminiscent of Yu-Shiba-Rosinov states \cite{YU_65,Shiba_1968,Rusinov_1969}. 
\begin{figure}
\begin{center}
\includegraphics[width=0.48\textwidth]{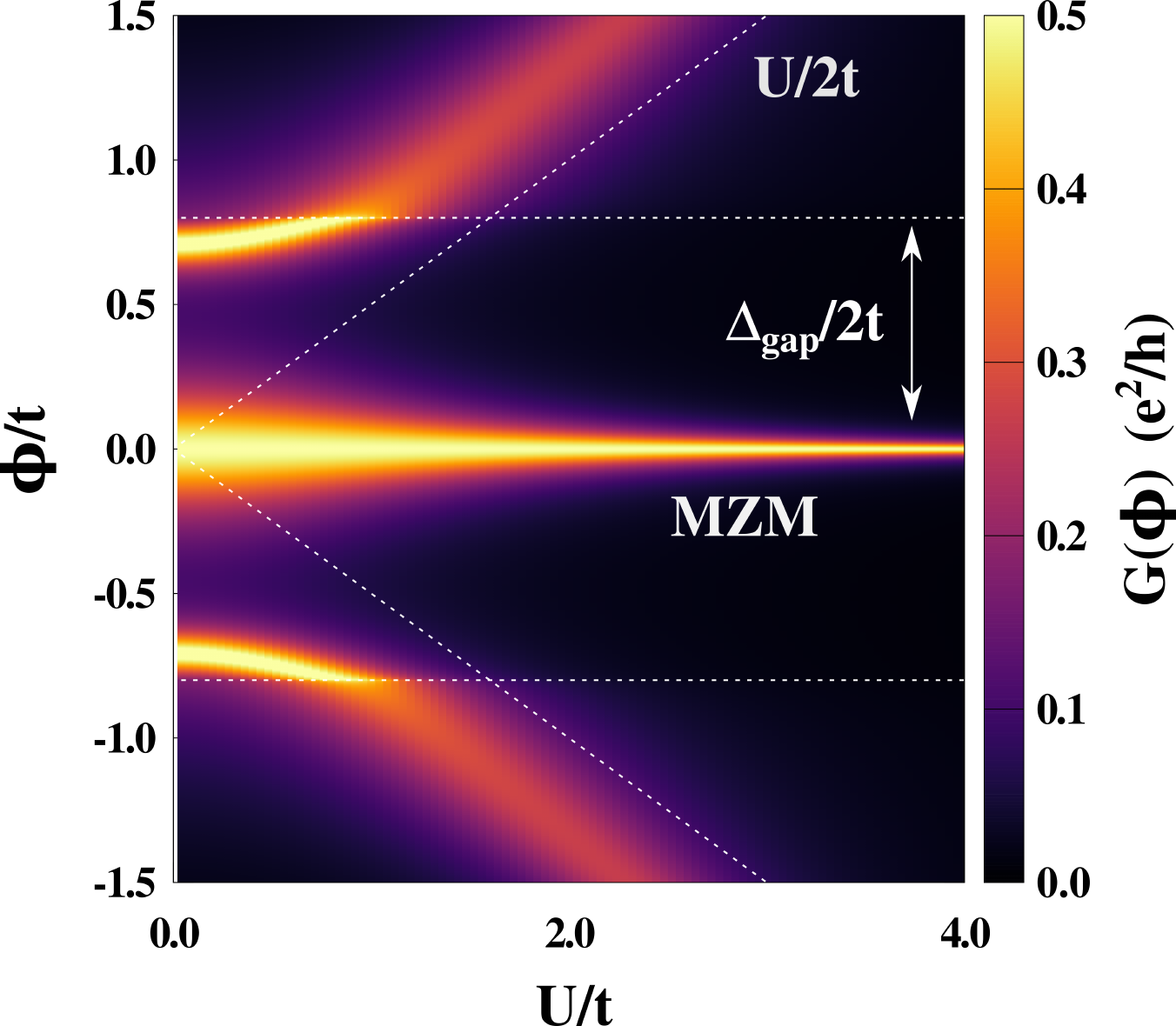}
\vspace{-0.45cm}
\caption{Influence of the interaction $U/t$ on the conductance $G(\phi)$ for $\mu/t=0.0$, $\Delta/t=V/t=0.4$ and $V_c/t=0.3$. Horizontal dashed lines show the width of the bulk superconducting gap $\Delta_{\text{gap}}$. }
\label{G_vs_U}
\end{center}
\end{figure} 
The effect of the interaction on the $G(\phi)$ is shown in Fig. \ref{G_vs_U}, where we report the evolution of the low-energy MZM and of the Yu-Shiba-Rosinov-like bound states as a function of $U/t$. Being non-topological, the latter features are strongly affected by the interaction, and indeed, as shown in Fig. \ref{G_vs_U}, above a certain value of $U/t$ they enter in the continuum of Cooper-pairs excitations of the SC. On the other hand, the contribution to the zero-bias conductance $G_0$ of the MZM is robust and persist for any value of $U/t$. The interaction renormalizes the coupling (\ref{H_T_Kitaev}) $V\to V\sqrt{Z}$ between the dot and the MZM by the quasiparticle renormalization factor $Z$, displayed in Fig. \ref{zeta_Uvsmu}, and enhances the curvature of the conductance close to the zero-bias anomaly (\ref{low_bias_current}).

\begin{figure}
\begin{center}
\includegraphics[width=0.48\textwidth]{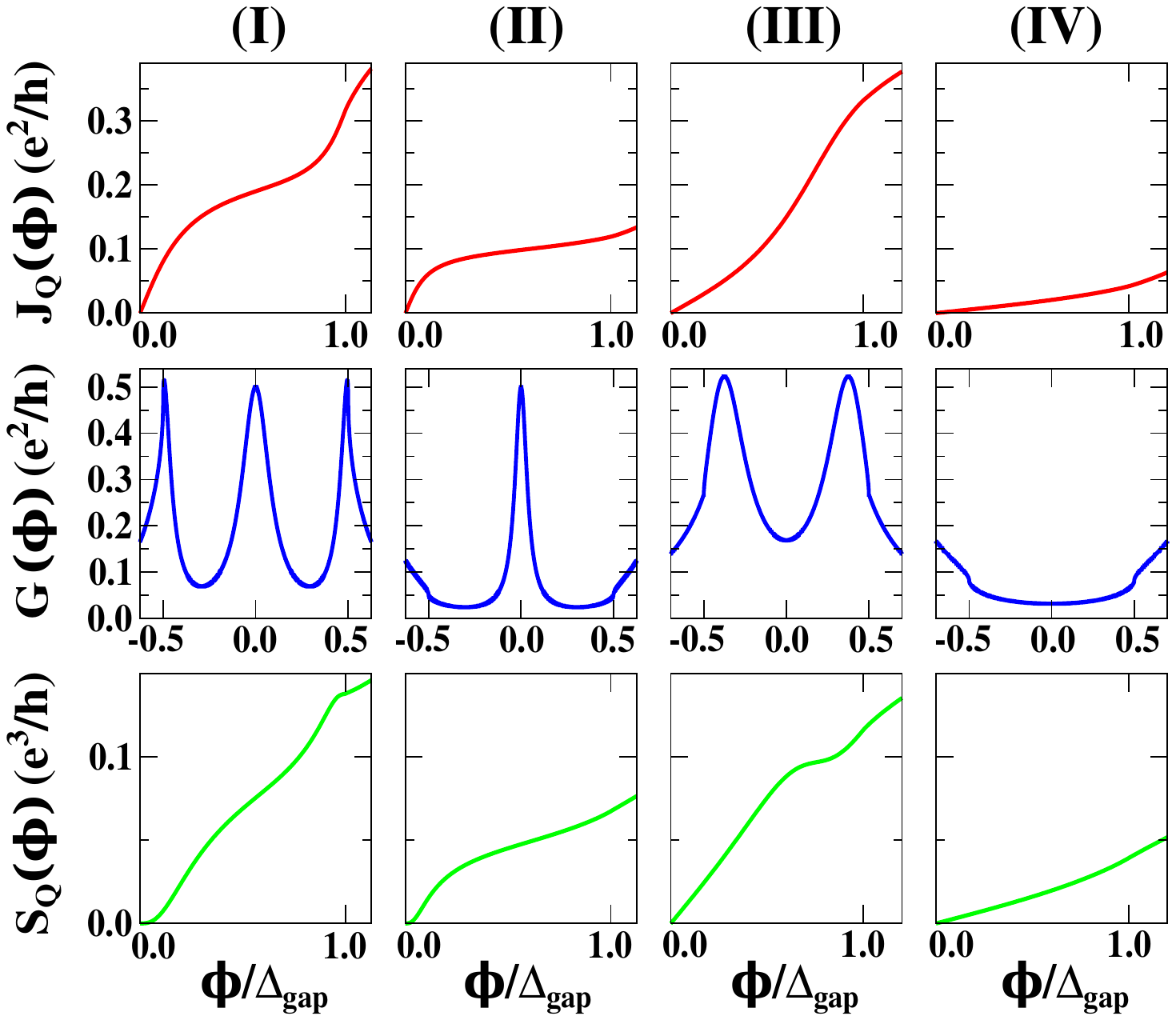}
\vspace{-0.5cm}
\caption{ From top to bottom current $J(\phi)$, conductance $G(\phi)$ and shot noise $S_Q(\phi)$  as a function of the bias $\phi/\Delta_{\text{gap}}$, where $\Delta_{\text{gap}}$ is the bulk superconducting gap. Left side panels (I) and (II) describe different regions in the Topological SC phase: parameters are $U/t=0.8,\,2.0$, $V/t=0.4$, $V_c/t=0.3$ and $\mu/t=0.0$ with bulk superconducting gap $\Delta_{\text{gap}}/t=1.6$. Right side panels (III) and (IV) describe different regions in the Trivial SC phase: parameters are $U/t=0.8,\,2.0$, $V/t=0.4$, $V_c/t=0.3$ and $\mu/t=2.5$ with bulk superconducting gap $\Delta_{\text{gap}}/t=1.0$. In the topological phase (I) and (II) the zero-bias anomaly $e^2/2h$ shows the presence of a MZM, that is absent in the trivial-SC, regions (III) and (IV).}
\label{JG_phases}
\end{center}
\end{figure}  
In order to have a complete characterization of the junction we compute the shot noise $S_Q$ that at zero temperature reads:
\bal
\label{shot_noise_charge}
S&_Q=\frac{2\pi e^3}{h}\frac{\pi}{2}\int d\epsilon\bar{\rho}^2(\epsilon)\frac{f_L(\epsilon)-f_R(\epsilon)}{2}\\
&\Tr\Big[\left(\hat{\mathbf{T}}^R_{\eta_\uparrow}(\epsilon)+\hat{\mathbf{T}}^A_{\eta_\uparrow}(\epsilon)\right)\cdot\left(\hat{\mathbf{T}}^R_{\eta_\uparrow}(\epsilon)+\hat{\mathbf{T}}^A_{\eta_\uparrow}(\epsilon)\right)\Big],
\eal
for more details we refer the interested reader to the supplementary material \cite{supplementary}.
Analogously to the previous case, we perform the average over $\downarrow$ configurations
\be
\label{noise_spin_av}
\langle S_Q\rangle=\sum_{n^f_\downarrow=0,1}p(n^f_\downarrow) S_Q(n^f_\downarrow).
\ee
A complete characterization of the low-energy transport properties is given in Fig. \ref{JG_phases}, where we plot the current $J_Q$, the corresponding charge-conductance $G(\phi)$ and its fluctuations $S_Q(\phi)$ as a function of the applied bias. We notice that dependently on the region of the Kitaev phase diagram \ref{zeta_Uvsmu} we predict different low-energy response. In particular behavior (I) and (II) denote the presence of a MZM, while (III) and (IV) characterize the non-topological regime. 
Differently from (I) and (IV), regions (II) and (III) present additional in-gap bound states distinguished by sharp peaks in $G(\phi)$ away from the zero-bias anomaly.
 \begin{figure}
\begin{center}
\includegraphics[width=0.47\textwidth]{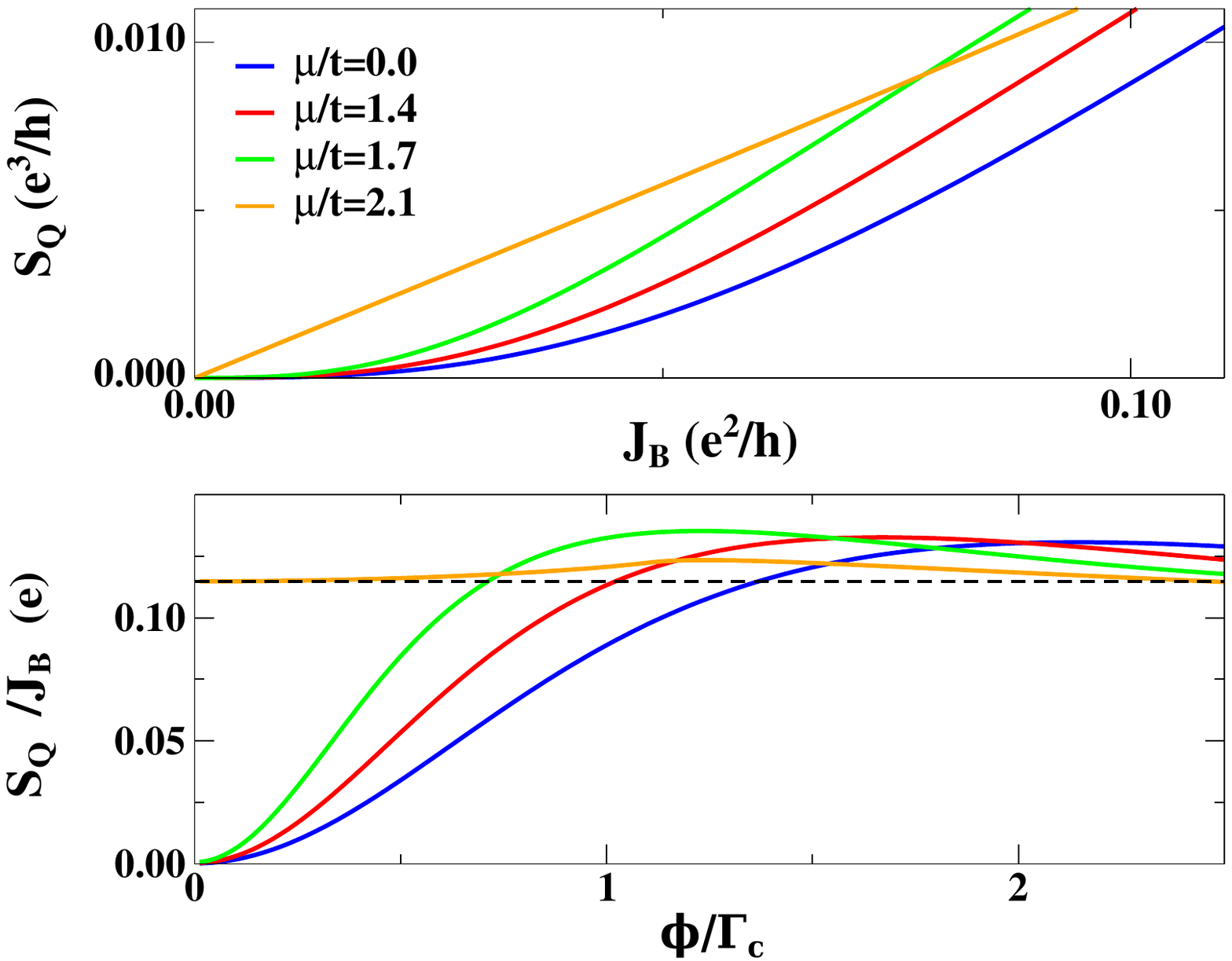}
\vspace{-0.1cm}
\caption{ Top Panel: shot noise $\langle S_{Q}\rangle$ as a function of the backscattering current $J_B$. Bottom panel: Ratio $\langle S_Q\rangle/J_B$ as a function of the bias $\phi$ measured in units of the hybridization with the metallic contacts $\Gamma_c/t=2\pi\overline{\rho}(0)V^2_c/t=0.18$. Dashed black line corresponds to Eq. (\ref{shot_noise_Coulomb_blockade}).
Different lines represent different values of the chemical potential of the Kitaev chain: $\mu/t=0.0,\,1.4,\,1.7$ and $2.1$. The other parameters are $V/t=\Delta/t=0.4$, $V_c/t=0.3$ and $U/\Gamma_c=5.\bar{5}$.}
\label{Fano_F}
\end{center}
\end{figure}

We observe that an additional signature of the MZM is given by the low-bias behavior of the shot noise $S_Q(\phi)$, which is shown in the bottom panel of Fig. \ref{JG_phases}. 
Indeed, in the topological regime, for small bias, the shot noise goes like: 
\be
\label{low_bias_SN}
\langle S_Q\rangle\simeq\frac{e^3}{h}\frac{\phi^3}{24\left(\Gamma_cZ\right)^2}\left[1-\frac{3}{10}\left(\frac{\phi}{\Gamma_cZ}\right)^2\right],
\ee
while it becomes linear in the non-topological region, $S_Q \propto\phi$ for $|\mu|>2t$.
The evaluation of the shot noise allows to compute the Fano factor
\be
\label{Fano_factor}
F=\frac{S_Q}{J_{B}}\Big|_{\phi=0}=q_{\text{e}}
\ee
 which determines the charge of the elementary carriers \cite{de-Picciotto1997}. In Eq. (\ref{Fano_factor}) we have introduced the backscattering current, defined as the deviation from the unitary transmission through the junction \cite{Eran_Sela_2006}:
 \be
 \label{unitary_T}
 J_{B}=\frac{e^2}{h}\phi-\langle\langle J_{Q}\rangle\rangle.
 \ee
As a consequence of the small bias behavior of Eqs. (\ref{low_bias_current}) and (\ref{low_bias_SN}), the topological regime $|\mu|<2t$ is characterized by a vanishing Fano factor $F=0$, independently from the value of the interaction $U/t$, as shown in the bottom panel of Fig. \ref{Fano_F}. On the other hand, in the non-topological region $F$ is a function of $U/t$ which becomes equal to $1$ in the non-interacting limit $U/t\to0$.  In particular for $|\mu|>2t$ we find:
\be
\label{shot_noise_Coulomb_blockade}
F=\frac{\left(2\Gamma_c/U\right)^2}{1+\left(2\Gamma_c/U\right)^2}.
\ee

Therefore, experimental measurements of the shot noise give additional informations complementary to those attainable by studying the characteristic zero-bias conductance $e^2/2h$. Combined measurements of the conductance and the shot noise in the experimental set-up presented in Fig. \ref{system3d} allow to detect the topological properties of the superconducting wire and to distinguish the low-energy contribution of a MZM from other possible sources of zero-bias anomaly. 
 We argue that the predicted behavior of the conductance and the shot noise persists even for a more realistic model Hamiltonian, that presents a non-vanishing tunnel-coupling with the spin $\downarrow$ fermionic operator in the quantum dot \cite{Rosa_2013,Cheng_2014,Weymann_2017,Weymann_2018_SciRep}. However, a detailed analysis of this problem is left to future investigations.
  
\paragraph*{Conclusions.}
The present results show that transport measurements give a detailed characterization of the topological phase diagram of real materials and reveal MZM in nano-wires. The presence of a MZM is outlined by a fractional zero-bias conductance $e^2/2h$ that, we have shown, is robust against the dot interaction. Additionally, for small values of the on-site repulsion, we find in-gap bound states that represent the only low-energy feature in the topologically trivial region of the phase diagram in Fig. \ref{zeta_Uvsmu}. 
Furthermore, we find that the topological regime is characterized by a vanishing Fano factor induced by the tunnel-coupling with the MZM at the edge of the superconducting wire. 
Our analysis gives a complete characterization of charge transport measurements that can experimentally detect the presence of MZM on the edge of real materials and, indirectly, allows to reconstruct their topological phase diagram.

\paragraph*{Acknowledgements.}
This work has been supported by the European Union under H2020 Framework Programs, ERC Advanced Grant No. 692670 ``FIRSTORM''. 
We are grateful to Michele Fabrizio, Domenico Giuliano and Roberto Raimondi for discussions and comments on the manuscript. We thank Shankar Ganesh and Joseph Maciejko for useful correspondence at the early stage of this work. 

\bibliographystyle{apsrev}
\bibliography{mybiblio}

\begin{thebibliography}{69}
\bibitem[{com()}]{comment_Shankar}
\bibinfo{note}{In an interesting recent paper \cite{Shankar_MAI} the authors
  used the $\mathbb{Z}_2$ slave-spin mapping to rewrite the original
  interacting Hamiltonian in a quadratic form. The same Hamiltonian can be
  obtained by performing the simple transformation in Eq. (\ref{GAUGE_xi})
  without introducing any additional degree of freedom.}
\end{thebibliography}
\end{document}


\title{Supplemental material for "Probing Majorana edge states by measuring transport through an interacting magnetic impurity"}
\author{Daniele Guerci}
\affiliation{International School for
  Advanced Studies (SISSA), Via Bonomea
  265, I-34136 Trieste, Italy} 
\author{Andrea Nava}
\affiliation{International School for
  Advanced Studies (SISSA), Via Bonomea
  265, I-34136 Trieste, Italy} 

\date{\today} 
\pacs{}

\maketitle

In this supplemental material we derive the main equations and results presented in the manuscript. In the first part of the appendix \ref{Tmat} we present the evaluation of the transfer matrix $\hat{\mathbf{T}}_{\eta_\uparrow}$, which allows to compute the charge current and the shot noise.  
 In the second part \ref{BdG} of the supplemental material, instead, we compute the wave functions of a magnetic impurity coupled to two Kitaev chains, that host a zero energy Majorana. The latter quantity allow to characterize the different scattering processes occurring at the junction and contributing to the charge current between the metallic contacts. 
\section{Charge transport within Nambu-Keldysh formalism}
\label{Tmat}
The model Hamiltonian we use to describe a magnetic impurity coupled to two metallic contacts and a p-wave superconductor reads:
\bal
\label{model_junction_SS_B}
H^{*}&=H_{C}+H_{K}+\frac{1}{2}\sum_{\alpha=L,R}\left(\vec{\eta}^\dagger_\uparrow\cdot\hat{V}^\alpha_c\cdot\vec{c}_{1\alpha}+\text{H.c.}\right)\\
&+\frac{1}{2}\left(\vec{\eta}^\dagger_\uparrow\cdot\hat{V}\cdot\vec{c}_1+\text{H.c.}\right)-\frac{1}{2}\vec{\eta}^\dagger_\uparrow\cdot\frac{U}{2}\sigma^z\cdot\vec{\eta}_\uparrow
\eal
where in the Nambu representation $\vec{\psi}=(\psi,\,\,\psi^\dagger)^T$, $\hat{V}$ is the hybridization matrix between the dot and the $1$st site of the Kitaev chain:
\be
  \label{V_hopping}
  \hat{V}=iV\left(\begin{array}{cc}1 & 1 \\1 & 1\end{array}\right)
\ee
and $\hat{V}_c$ couples the metallic contacts to the dot
\be
  \label{V_metal}
  \hat{V}^\alpha_{c}=\frac{V^\alpha_c}{2}\left(\begin{array}{cc}1+q^d_\downarrow & -(1-q^d_\downarrow) \\(1-q^d_\downarrow) & -(1+q^d_\downarrow)\end{array}\right).
\ee
In Hamiltonian (\ref{model_junction_SS_B}) $H_c$ and $H_K$ are semi-infinite $1-$D chains describing the metallic contacts and the Kitaev chain.
We remind that the number of $\downarrow$ dot fermion is conserved, $[d_{\downarrow}^{\dagger}d_{\downarrow},H]=0$ and  $q^d_\downarrow=(1-2d_{\downarrow}^{\dagger}d_{\downarrow})$ is an effective
$\mathbb{Z}_2$ variable $q^d_\downarrow=(-1,1) $. As already observed in the manuscript this property makes the model exactly solvable, similarly to the Falicov-Kimball model (FKM) \cite{Falicov_1969} the $\downarrow$ configuration is obtained by minimizing the ground-state energy of the $\uparrow$ degrees of freedom. 
\begin{table}
\caption{The table below summarizes the different Green's function involved in the evaluation of the charge transport properties of the model (\ref{model_junction_SS_B}).}
\vspace{0.5cm}
\centering
\begin{tabular}{ | m{11cm}| m{3cm} | } 
\hline
 \textbf{Green's functions of the junction}& \\
\hline
 $x, x^\prime$ infinite chain lead $\alpha$ & $\hat{G}_{xx^\prime\alpha\alpha}$ \\
\hline
  $x, x^\prime$ semi-infinite chain lead $\alpha$  & $\hat{\overline{G}}_{xx^\prime\alpha\alpha}$ \\
\hline
  $x, x^\prime$ semi-infinite chain hybridized with the impurity site leads $\alpha,\beta$  & $\hat{\overline{\mathbf{G}}}_{xx^\prime\alpha\beta}$ \\
\hline
 bare $\eta_\uparrow$  & $\hat{G}_{\eta_\uparrow}$ \\
 \hline
 dressed $\eta_\uparrow$  & $\hat{\mathbf{G}}_{\eta_\uparrow}$ \\
 \hline
 mixing between the impurity and the leads $\eta_\uparrow-1\alpha$ and $1\alpha-\eta_\uparrow$ & $\hat{G}_{1\alpha\eta_\uparrow},\,\hat{G}_{\eta_\uparrow1\alpha}$ \\
 \hline
\end{tabular}
\label{Tab_GF}  
\end{table}
Within the Nambu formalism we define the Keldysh Green's functions of a fermionic operator $\psi_\alpha$ as:
\be
\label{Keldysh_GF}
\hat{G}_{\alpha\beta}(t_s,t^{\prime}_{s^\prime})=-i\left\langle T_C\left(\begin{array}{cc} \psi_\alpha(t_s)\psi^\dagger_\beta(t^\prime_{s^\prime}) &  \psi_\alpha(t_s)\psi_\beta(t^\prime_{s^\prime})  \\
\psi^\dagger_\alpha(t_s)\psi^\dagger_\beta(t^\prime_{s^\prime}) & \psi^\dagger_\alpha(t_s)\psi_\beta(t^\prime_{s^\prime})
\end{array}\right)\right\rangle,
\ee   
where $T_C$ is the contour ordering and $s,s^\prime=\pm$, where $-$ and $+$ are the forward and backward branches of the Keldysh contour, respectively.  Notice well, in the following we refer to the lesser Green's function as $G^<(t,t')=G(t_-,t'_+)$ and the greater one as $G^>(t,t')=G(t_+,t'_-)$. The hybridization between the dot and the leads introduces several boundary Green's functions that are summarized in the table \ref{Tab_GF}. We remind that Green's functions are $4\times4$ matrices in the Nambu-Keldysh space.

By performing perturbation theory in the tunnel-coupling between the leads and the impurity we obtain the following Dyson's equation
\be
\label{G_eta}
\hat{\mathbf{G}}_{\eta_\uparrow}=\hat{G}_{\eta_\uparrow}+\hat{G}_{\eta_\uparrow}\bullet\hat{\Sigma}_{\eta_\uparrow}\bullet\hat{\mathbf{G}}_{\eta_\uparrow},
\ee
where $\bullet$ is the convolution
\be
\label{convolution}
A\bullet B=\int dt_1 A(t,t_1)B(t_1,t^\prime),
\ee
 and
\be
\label{eta_self}
\hat{\Sigma}_{\eta_\uparrow}=\sum_{\alpha=L,R}\hat{V}^\alpha_c\cdot\hat{\overline{G}}_{11\alpha\alpha}\cdot\left(\hat{V}^\alpha_c\right)^\dagger+\hat{V}\cdot\hat{\overline{G}}_{11}\cdot\hat{V}^\dagger,
\ee  
 with $\hat{\overline{G}}_{11\alpha\alpha}$ boundary Green's function of the metallic lead $\alpha$ and $\hat{\overline{G}}_{11}$ boundary Green's function of the superconductive chain, we refer to section \ref{BG_leads} for more details. After straightforward calculations we find that:
\be
\label{mixed_G}
\hat{G}_{1a\eta_\uparrow}=\hat{\overline{G}}_{11aa}\bullet\hat{V}^\dagger_a\cdot\hat{\mathbf{G}}_{\eta_\uparrow},\quad\hat{G}_{\eta_\uparrow1\alpha}=\hat{\mathbf{G}}_{\eta_\uparrow}\cdot\hat{V}_a\bullet\hat{\overline{G}}_{11aa}.
\ee 
where $a$ refers to one of the three leads connected to the impurity. Finally, the hybridization with the dot induces a direct coupling between different leads:
\be
\label{coupling_leads}
\hat{\overline{\mathbf{G}}}_{11ab}=\delta_{ab}\hat{\overline{G}}_{11aa}+\hat{\overline{G}}_{11aa}\bullet\hat{\mathbf{T}}^{ab}_{\eta_\uparrow}\bullet\hat{\overline{G}}_{11bb}
\ee
where the indices $a,b$ refer to the metallic leads as well as the Kitaev chain and we have introduced the transfer matrix:
\be
\label{transfer_matrix}
\hat{\mathbf{T}}^{ab}_{\eta_\uparrow}=\hat{V}^\dagger_a\cdot\hat{\mathbf{G}}_{\eta_\uparrow}\cdot \hat{V}_b.
\ee
In particular transport across the metallic contacts involves $\hat{\mathbf{T}}^{\alpha\beta}_{\eta_\uparrow}$ with $\alpha,\beta=L,R$. From now on we consider symmetric metallic leads, i.e. $\hat{V}^L_c=\hat{V}^R_c=\hat{V}_c$, such that $\hat{\mathbf{T}}^{\alpha\beta}_{\eta_\uparrow}=\hat{\mathbf{T}}_{\eta_\uparrow}$ does not depend on $\alpha$ and $\beta$.

\subsection{Charge current}

The current operator for the metallic lead $\alpha=L,R$ is:
\be
\label{lead_alpha_current}
J_\alpha=-i\frac{V_c}{2}\left(\gamma_{1\alpha}\gamma^\eta_\uparrow+q^d_\downarrow\xi_{1\alpha}\xi^\eta_{\uparrow}\right).
\ee
After straightforward calculations:
\be
\langle J_{\alpha}\rangle=\frac{\text{Tr}\left[\sigma^z\cdot\hat{G}^{<}_{1\alpha\eta_\uparrow}(t,t)\cdot\hat{V}_{c}\right]-\text{Tr}\left[\sigma^z\cdot\hat{V}^\dagger_{c}\cdot\hat{G}^{<}_{\eta_\uparrow1\alpha}(t,t)\right]}{2}.
\ee
The charge current, $J_Q=(J_L-J_R)/2$, across the junction reads
\bal
\label{current_1}
\langle J_{Q}\rangle&=\sum_{\alpha=L,R}\text{sign}(\alpha)\int\frac{d\omega}{2\pi}\frac{\text{Tr}\left[\sigma^z\cdot\hat{\overline{G}}^R_{11\alpha\alpha}(\omega)\cdot\hat{\mathbf{T}}^<_{\eta_\uparrow}(\omega)\right]-\text{Tr}\left[\sigma^z\cdot\hat{\mathbf{T}}^{<}_{\eta_\uparrow}(\omega)\cdot\hat{\overline{G}}^{A}_{11\alpha\alpha}(\omega)\right]}{4}\\
&+\sum_{\alpha=L,R}\text{sign}(\alpha)\int\frac{d\omega}{2\pi}\frac{\text{Tr}\left[\sigma^z\cdot\hat{\overline{G}}^{<}_{11\alpha\alpha}(\omega)\cdot\hat{\mathbf{T}}^A_{\eta_\uparrow}(\omega)\right]-\text{Tr}\left[\sigma^z\cdot\hat{\mathbf{T}}^{R}_{\eta_\uparrow}(\omega)\cdot\hat{\overline{G}}^<_{11\alpha\alpha}(\omega)\right]}{4},
\eal
where $\text{sign}(L)=+1$ and $\text{sign}(R)=-1$. 
We notice that $L$ and $R$ leads are characterized by the same hybridization matrix $\hat{V}_c$ as well as the same spectral properties. Therefore, the first contribution to the current in Eq. (\ref{current_1}) vanishes and:
 \bal
\langle J_{Q}\rangle&=\int\frac{d\omega}{2\pi}\frac{\text{Tr}\left[\sigma^z\cdot\left(\hat{\overline{G}}^{<}_{11LL}(\omega)-\hat{\overline{G}}^{<}_{11RR}(\omega)\right)\cdot\hat{T}^A(\omega)\right]-\text{Tr}\left[\sigma^z\cdot\hat{T}^{R}(\omega)\cdot\left(\hat{\overline{G}}^{<}_{11LL}(\omega)-\hat{\overline{G}}^{<}_{11RR}(\omega)\right)\right]}{4},
\eal
where
\be
\hat{\overline{G}}^{<}_{11LL}(\omega)-\hat{\overline{G}}^{<}_{11RR}(\omega)=2\pi i\overline{\rho}(\omega)\left(f_{L}(\omega)-f_{R}(\omega)\right)\sigma^z,
\ee
and $\overline{\rho}(\omega)$ is the boundary spectral function of the metallic leads (\ref{Semicircular_DOS}). By rescaling for $-2\pi e^2/h$ we obtain
\be
\label{final_current}
\langle J_Q\rangle=\frac{\pi e^2}{h}\int d\epsilon \overline{\rho}(\epsilon)\left(f_L(\epsilon)-f_R(\epsilon)\right)\text{Im}\Tr\left(\hat{\mathbf{T}}^A_{\eta_\uparrow}(\epsilon)\right).
\ee
 Moreover, we notice that 
\be
\label{impurity_transfer}
\hat{\mathbf{T}}^{A}(\epsilon)=\hat{V}^\dagger_{c}\cdot \hat{\mathbf{G}}^{A}_{\eta_\uparrow}(\epsilon)\cdot \hat{V}_{c},
\ee
and 
\be
\label{GF_eta}
\left[\hat{\mathbf{G}}^{A}_{\eta_\uparrow}(\epsilon) \right]^{-1}=\epsilon\mathbf{1}+\frac{U}{2}\sigma^z-\sum_{\alpha=L,R}\hat{V}_c\cdot\hat{\overline{G}}^{A}_{11\alpha\alpha}(\omega)\cdot\hat{V}^\dagger_c-\hat{V}\cdot\hat{\overline{G}}^{A}_{11}(\omega)\cdot\hat{V}^\dagger,
\ee
with $\hat{\overline{G}}^{A}_{11\alpha\alpha}(\omega)$ boundary Green's function of the metallic contacts (\ref{Boundary_RA_expression_met_ex}) and $\hat{\overline{G}}^{A}_{11}(\omega)$ of the Kitaev chain Eqs. (\ref{SC_local_GR}), (\ref{SC_nn_GR}) and (\ref{Boundary_RA_expression_met}).  
The value of the current is obtained by averaging Eq. (\ref{final_current}) over the spin $\downarrow$ configurations:
\be
\label{charge_spin_av}
\langle\langle J_Q\rangle\rangle=\sum_{n^f_\downarrow=0,1}p(n^f_\downarrow)\langle J_Q(n^f_\downarrow)\rangle
\ee
where in the absence of any gate potential or Zeeman field on the quantum dot $p(0)=p(1)=1/2$.

\subsection{Shot noise}

The correlation function between currents $J_\alpha$ and $J_\beta$ reads:
\be
S_{\alpha\beta}(t,t^\prime)=\langle T_C\left(\delta J_\alpha(t)\delta J_\beta(t^\prime)\right)\rangle
\ee
where $T_C$ is the time-ordering operator on the Keldysh contour, $\delta J_\alpha = J_\alpha - \langle J_\alpha\rangle$. In the following we evaluate the $T_C$ ordered $S_{\alpha\beta}$, where usual perturbation theory can be applied, and then we take the lesser and greater components to compute $S_{Q}(t,t^\prime)$. 
After straightforward calculations we arrive at the following expression:
\bal
\label{CC-Tt}
S_{\alpha\beta}(t,t^\prime)&=\frac{1}{2}\Big[\Tr\left(\sigma^z\cdot\hat{\mathbf{T}}_{\eta_\uparrow}(t,t^\prime)\cdot\sigma^z\cdot\hat{\overline{\mathbf{G}}}_{11\beta\alpha}(t^\prime,t)\right)+\Tr\left(\sigma^z\cdot\hat{\overline{\mathbf{G}}}_{11\alpha\beta}(t,t^\prime)\cdot\sigma^z\cdot\hat{\mathbf{T}}_{\eta_\uparrow}(t^\prime,t)\right)\\
&-\Tr\left(\sigma^z\cdot\hat{V}^\dagger_c\cdot\hat{G}_{\eta1\beta}(t,t^\prime)\cdot\sigma^z\cdot\hat{V}^\dagger_c\cdot\hat{G}_{\eta1\alpha}(t^\prime,t)\right)-
\Tr\left(\sigma^z\cdot\hat{G}_{1\alpha\eta}(t,t^\prime)\cdot\hat{V}_c\cdot\sigma^z\cdot\hat{G}_{1\beta\eta}(t^\prime,t)\cdot\hat{V}_c\right)\Big].
\eal
We are interested in average values of the form
\bal
\langle\{\delta J_\alpha(t),\delta J_\beta(t^\prime)\}\rangle&=\langle T_C\left(\delta J_\alpha(t_-)\delta J_\beta(t^\prime_+)\right)\rangle+\langle T_C\left(\delta J_\alpha(t_+)\delta J_\beta(t^\prime_-)\right)\rangle\\
&=S^<_{\alpha\beta}(t,t^\prime)+S^>_{\alpha\beta}(t,t^\prime),
\eal
where:
\bal
\label{CC-Tt_Lesser}
S^{<}_{\alpha\beta}(t,t^\prime)&=\frac{1}{2}\Big[\Tr\left(\sigma^z\cdot\hat{\mathbf{T}}^<_{\eta_\uparrow}(t,t^\prime)\cdot\sigma^z\cdot\hat{\overline{\mathbf{G}}}^>_{11\beta\alpha}(t^\prime,t)\right)+\Tr\left(\sigma^z\cdot\hat{\overline{\mathbf{G}}}^<_{11\alpha\beta}(t,t^\prime)\cdot\sigma^z\cdot\hat{\mathbf{T}}^>_{\eta_\uparrow}(t^\prime,t)\right)\\
&-\Tr\left(\sigma^z\cdot\hat{V}^\dagger_c\cdot\hat{G}^<_{\eta1\beta}(t,t^\prime)\cdot\sigma^z\cdot\hat{V}^\dagger_c\cdot\hat{G}^>_{\eta1\alpha}(t^\prime,t)\right)-
\Tr\left(\sigma^z\cdot\hat{G}^<_{1\alpha\eta}(t,t^\prime)\cdot\hat{V}_c\cdot\sigma^z\cdot\hat{G}^>_{1\beta\eta}(t^\prime,t)\cdot\hat{V}_c\right)\Big],
\eal
and 
\bal
\label{CC-Tt_Greater}
S^{>}_{\alpha\beta}(t,t^\prime)&=\frac{1}{2}\Big[\Tr\left(\sigma^z\cdot\hat{\mathbf{T}}^>_{\eta_\uparrow}(t,t^\prime)\cdot\sigma^z\cdot\hat{\overline{\mathbf{G}}}^<_{11\beta\alpha}(t^\prime,t)\right)+\Tr\left(\sigma^z\cdot\hat{\overline{\mathbf{G}}}^>_{11\alpha\beta}(t,t^\prime)\cdot\sigma^z\cdot\hat{\mathbf{T}}^<_{\eta_\uparrow}(t^\prime,t)\right)\\
&-\Tr\left(\sigma^z\cdot\hat{V}^\dagger_c\cdot\hat{G}^>_{\eta1\beta}(t,t^\prime)\cdot\sigma^z\cdot\hat{V}^\dagger_c\cdot\hat{G}^<_{\eta1\alpha}(t^\prime,t)\right)-
\Tr\left(\sigma^z\cdot\hat{G}^>_{1\alpha\eta}(t,t^\prime)\cdot\hat{V}_c\cdot\sigma^z\cdot\hat{G}^<_{1\beta\eta}(t^\prime,t)\cdot\hat{V}_c\right)\Big].
\eal
In the steady-state regime we define the zero frequency limit of the current-current response spectrum as:
\be
\mathcal{P}_{\alpha\beta}=\lim_{\Omega\to0}\int dt e^{-i\Omega(t-t^\prime)}\frac{\langle\{\delta J_\alpha(t),\delta J_\beta(t^\prime)\}\rangle}{2}=\frac{S^{<}_{\alpha\beta}(\Omega=0)+S^{>}_{\alpha\beta}(\Omega=0)}{2},
\ee
where
\bal
\label{CC-Tt_Lesser_omega}
S^{<}_{\alpha\beta}(\Omega=0)&=\int\frac{d\omega}{4\pi}\Big[\Tr\left(\sigma^z\cdot\hat{\mathbf{T}}^<_{\eta_\uparrow}(\omega)\cdot\sigma^z\cdot\hat{\overline{\mathbf{G}}}^>_{11\beta\alpha}(\omega)\right)+\Tr\left(\sigma^z\cdot\hat{\overline{\mathbf{G}}}^<_{11\alpha\beta}(\omega)\cdot\sigma^z\cdot\hat{\mathbf{T}}^>_{\eta_\uparrow}(\omega)\right)\\
&-\Tr\left(\sigma^z\cdot\hat{V}^\dagger_c\cdot\hat{G}^<_{\eta1\beta}(\omega)\cdot\sigma^z\cdot\hat{V}^\dagger_c\cdot\hat{G}^>_{\eta1\alpha}(\omega)\right)-
\Tr\left(\sigma^z\cdot\hat{G}^<_{1\alpha\eta}(\omega)\cdot\hat{V}_c\cdot\sigma^z\cdot\hat{G}^>_{1\beta\eta}(\omega)\cdot\hat{V}_c\right)\Big],
\eal
and 
\bal
\label{CC-Tt_Greater_omega}
S^{>}_{\alpha\beta}(\Omega=0)&=\int\frac{d\omega}{4\pi}\Big[\Tr\left(\sigma^z\cdot\hat{\mathbf{T}}^>_{\eta_\uparrow}(\omega)\cdot\sigma^z\cdot\hat{\overline{\mathbf{G}}}^<_{11\beta\alpha}(\omega)\right)+\Tr\left(\sigma^z\cdot\hat{\overline{\mathbf{G}}}^>_{11\alpha\beta}(\omega)\cdot\sigma^z\cdot\hat{\mathbf{T}}^<_{\eta_\uparrow}(\omega)\right)\\
&-\Tr\left(\sigma^z\cdot\hat{V}^\dagger_c\cdot\hat{G}^>_{\eta1\beta}(\omega)\cdot\sigma^z\cdot\hat{V}^\dagger_c\cdot\hat{G}^<_{\eta1\alpha}(\omega)\right)-
\Tr\left(\sigma^z\cdot\hat{G}^>_{1\alpha\eta}(\omega)\cdot\hat{V}_c\cdot\sigma^z\cdot\hat{G}^<_{1\beta\eta}(\omega)\cdot\hat{V}_c\right)\Big].
\eal
We notice that 
\be
\label{symmetry}
S^>_{\alpha\beta}(0)=S^<_{\beta\alpha}(0)\implies\mathcal{P}_{LR}=\mathcal{P}_{RL},
\ee
and 
\be
\label{noise}
S_{Q}(\Omega=0)=\frac{\mathcal{P}_{LL}+\mathcal{P}_{RR}-2\mathcal{P}_{LR}}{4}.
\ee
By using Eqs. (\ref{G_eta}), (\ref{mixed_G}) and (\ref{coupling_leads})  we obtain:
\bal
\label{S_Q}
S_{Q}&=\frac{1}{16\pi}\int d\omega\Big[\Tr\left(\sigma^z\cdot\hat{\mathbf{T}}^<_{\eta_\uparrow}(\omega)\cdot\sigma^z\cdot\left(\hat{\overline{G}}^>_{11LL}(\omega)+\hat{\overline{G}}^>_{11RR}(\omega)\right)\right)+\Tr\left(\sigma^z\cdot\hat{\mathbf{T}}^>_{\eta_\uparrow}(\omega)\cdot\sigma^z\cdot\left(\hat{\overline{G}}^<_{11LL}(\omega)+\hat{\overline{G}}^<_{11RR}(\omega)\right)\right)\Big]\\
&+\frac{1}{16\pi}\int d\omega \Big[\Tr\left(\sigma^z\cdot\hat{\mathbf{T}}^R_{\eta_\uparrow}(\omega)\cdot\left(\hat{\overline{G}}^<_{11LL}(\omega)-\hat{\overline{G}}^<_{11RR}(\omega)\right)\cdot\sigma^z\cdot\hat{\mathbf{T}}^R_{\eta_\uparrow}(\omega)\cdot\left(\hat{\overline{G}}^>_{11RR}(\omega)-\hat{\overline{G}}^>_{11LL}(\omega)\right)\right)\\
&+\Tr\left(\sigma^z\cdot\hat{\mathbf{T}}^A_{\eta_\uparrow}(\omega)\cdot\left(\hat{\overline{G}}^<_{11LL}(\omega)-\hat{\overline{G}}^<_{11RR}(\omega)\right)\cdot\sigma^z\cdot\hat{\mathbf{T}}^A_{\eta_\uparrow}(\omega)\cdot\left(\hat{\overline{G}}^>_{11RR}(\omega)-\hat{\overline{G}}^>_{11LL}(\omega)\right)\right)\Big].
\eal
By using Eqs. (\ref{Boundary_RA_expression_met_ex}) and (\ref{BC_xx_Lesser_Greater}) we obtain
\bal
\label{differences}
&\hat{\overline{G}}^<_{11LL}(\omega)-\hat{\overline{G}}^<_{11RR}(\omega)=2i\pi\bar{\rho}(\omega)(f_L(\omega)-f_R(\omega))\sigma^z,\\
&\hat{\overline{G}}^>_{11RR}(\omega)-\hat{\overline{G}}^>_{11LL}(\omega)=-2i\pi\bar{\rho}(\omega)(f_L(\omega)-f_R(\omega))\sigma^z,\\
&\hat{\overline{G}}^<_{11LL}(\omega)+\hat{\overline{G}}^<_{11RR}(\omega)=2i\pi\bar{\rho}(\omega)(f_L(\omega)+f_R(\omega))\sigma^0,\\
&\hat{\overline{G}}^>_{11LL}(\omega)+\hat{\overline{G}}^>_{11RR}(\omega)=-2i\pi\bar{\rho}(\omega)(2-f_L(\omega)-f_R(\omega))\sigma^0.
\eal
Finally, the expression of the white-noise component of $J_Q$ fluctuations reads:
\bal
\label{white_noise}
S_Q=&\pi\int d\omega\bar{\rho}^2(\omega)\Big[\frac{2-f_L(\omega)-f_R(\omega)}{2}\frac{f_L(\omega)+f_R(\omega)}{2}\left(\Tr(\hat{\mathbf{T}}^R_{\eta_\uparrow}(\omega)\cdot\hat{\mathbf{T}}^A_{\eta_\uparrow}(\omega))+\Tr(\hat{\mathbf{T}}^A_{\eta_\uparrow}(\omega)\cdot\hat{\mathbf{T}}^R_{\eta_\uparrow}(\omega))\right)\\
&+\frac{f_L(\omega)-f_R(\omega)}{2}\frac{f_L(\omega)-f_R(\omega)}{2}\left(\Tr(\hat{\mathbf{T}}^R_{\eta_\uparrow}(\omega)\cdot\hat{\mathbf{T}}^R_{\eta_\uparrow}(\omega))+\Tr(\hat{\mathbf{T}}^A_{\eta_\uparrow}(\omega)\cdot\hat{\mathbf{T}}^A_{\eta_\uparrow}(\omega))\right)\Big].
\eal
Since we are interested in quantum-fluctuations we take the zero-temperature limit:
\be
\label{phase_space}
\frac{2-f_L(\omega)-f_R(\omega)}{2}\frac{f_L(\omega)+f_R(\omega)}{2}=\frac{f_L(\omega)-f_R(\omega)}{2}\frac{f_L(\omega)-f_R(\omega)}{2}=\frac{f_L(\omega)-f_R(\omega)}{4},
\ee  
\be
\label{zerotemperature_shotnoise}
S_Q=\frac{2\pi e^3}{h}\frac{\pi}{4}\int_{-\phi/2}^{\phi/2} d\omega\overline{\rho}^2(\omega)\Tr\Big[\left(\hat{\mathbf{T}}^R_{\eta_\uparrow}(\omega)+\hat{\mathbf{T}}^A_{\eta_\uparrow}(\omega)\right)\cdot\left(\hat{\mathbf{T}}^R_{\eta_\uparrow}(\omega)+\hat{\mathbf{T}}^A_{\eta_\uparrow}(\omega)\right)\Big],
\ee
where $2\pi e^3/h$ is the rescaling factor, $\overline{\rho}(\omega)$ is the boundary DOS of the metallic chain (\ref{Semicircular_DOS}) and $\hat{\mathbf{T}}^{R/A}_{\eta_\uparrow}$ is given in Eqs. (\ref{impurity_transfer}) and (\ref{GF_eta}). The shot noise is obtained by averaging over the spin $\downarrow$ 
\be
\langle S_{Q}\rangle=\sum_{n_\downarrow=0,1}p(n_\downarrow)S_{Q}(n_\downarrow),
\ee
where $p(0)=p(1)=1/2$ at half-filling.

\subsection{Boundary Green's functions of the leads}
\label{BG_leads}
In this Appendix, we provide a derivation of the retarded/advanced and lesser/greater Green's function used in the manuscript, which describes quasiparticle excitations at the boundary of the semi-infinite leads \cite{Zazunov_2016,Zazunov_2017,Shankar_MAI}.

\paragraph{Metallic leads.}
\label{metallic_contacts}

In the Nambu formalism the $\alpha=L,R$ $k$-space Hamiltonian is:
\be
\label{PBC_model}
\mathcal{H}_{\alpha k}=\xi^{\alpha}_{k}\sigma^z,
\ee
where $\xi^{\alpha}_k=\epsilon_k-\mu_\alpha$, and $\epsilon_k=-2t\cos k$. 
In the following we will firstly compute the retarded $(R)$ and advanced $(A)$ boundary Green's function and then the lesser $(<)$ and greater $(>)$ ones.
To this aim we remind that the Green's function of a metallic chain with periodic boundary conditions reads:
\be
\label{metallic}
\hat{G}_{k\alpha\alpha}(z)=\frac{z}{z^2-\epsilon^2_k}\mathbf{1}+\frac{\epsilon_k}{z^2-\epsilon^2_k}\sigma^z
\ee
where $\mathbf{1}$ is the identity and $\sigma^z$ the third Pauli matrix, as usual the effect of the electrochemical potential $\mu_\alpha$ enters in statistical averages and does not influence the spectral properties of the metallic contacts. 

The $\hat{G}_{xx^\prime\alpha \alpha }$ Green's function is:
\bal
\label{metallic_real_space}
\hat{G}_{xx^\prime\alpha \alpha }(z)&=\sum_{s=\pm}\int_{0}^{\pi}\frac{dk}{2\pi}e^{isk(x-x^{\prime})}\hat{G}_{sk\alpha \alpha }(z),\\
&=\sum_{s=\pm}\int_{-1}^{1}\frac{dy}{2\pi\sqrt{1-y^2}}\left(-y+is\sqrt{1-y^2}\right)^{x-x^\prime}\left(\frac{z}{z^2-4t^2y^2}\mathbf{1}+\frac{2ty}{z^2-4t^2y^2}\sigma^z\right),
\eal
where we have performed the change of variable $z=-\cos k$. In order to compute the boundary Green's function of the semi-infinite metallic chain we need the local Green's function $\hat{G}_{xx\alpha\alpha}$: 
\be
\label{local_G}
\hat{G}_{xx\alpha \alpha }(z)=\frac{1}{z\sqrt{1-\left(2t/z\right)^2}}\mathbf{1},
\ee
and the nearest-neighbor ones 
\be
\label{nn_G}
\hat{G}_{x+1x\alpha \alpha }(z)=\hat{G}_{xx+1\alpha\alpha }(z)=\frac{1}{2t}\left(1-\frac{1}{\sqrt{1-\left(2t/z\right)^2}}\right)\sigma^z.
\ee
Moreover, we have: 
\be
\label{x_x_prime_Lesser}
\hat{G}^<_{xx^\prime\alpha\alpha}(\omega)=i\pi\rho(\omega)\sum_{s=\pm1}\left(\begin{array}{cc}f(\omega-\mu_\alpha)\left[-\frac{\omega}{2t}+is\sqrt{1-\left(\frac{\omega}{2t}\right)^2}\right]^{x-x^\prime} & 0 \\0 & f(\omega+\mu_\alpha)\left[\frac{\omega}{2t}+is\sqrt{1-\left(\frac{\omega}{2t}\right)^2}\right]^{x-x^\prime} \end{array}\right)
\ee
where $\rho(\omega)$ is the $1$-D density of states
\be
\label{1D_DOS}
\rho(\omega)=\frac{1}{2t\pi\sqrt{1-\left(\omega/2t\right)^2}}\theta\left(1-\frac{|\omega|}{2t}\right).
\ee

The boundary Green's function of a semi-infinite metallic chain located at $x > 0$ can be obtained from the "bulk" Green's function for the translationally invariant model (\ref{PBC_model}) by adding a local impurity of strength $\lambda$ at site $x = 0$, which results in the perturbation:
\be
\label{perturbation}
\Delta H= \lambda c^\dagger_0c_0.
\ee
By performing perturbation theory in (\ref{perturbation}) we obtain the following Dyson's equations for the boundary Green's functions:
\bal
\label{Dyson_boundary}
 &\hat{\overline{G}}_{xx}=\hat{G}_{xx}+\hat{G}_{x0}\bullet \lambda\sigma^z\hat{\overline{G}}_{0x},\\
 &\hat{\overline{G}}_{0x}=\hat{G}_{0x}+\hat{G}_{00} \bullet \lambda\sigma^z\hat{\overline{G}}_{0x},
 \eal
 where for simplicity we drop the index $\alpha$.
In the limit $\lambda\to\infty$, i.e., when one effectively cuts the wire into two semi-infinite pieces, Eq. (\ref{Dyson_boundary}) yields for the boundary Green's functions. Thus, by taking the lesser component of Eq. (\ref{Dyson_boundary}) and then the $\lambda\to\infty$ limit $\hat{\overline{G}}^<_{xx^\prime}$ reads:
\be
\label{Boundary_lesser_expression}
\hat{\overline{G}}^<_{xx^\prime}=\hat{G}^<_{xx^\prime}-\hat{G}^R_{x0}\bullet\left(\hat{G}^R_{00}\right)^{-1}\bullet\hat{G}^<_{0x^\prime}-\hat{G}^<_{x0}\bullet\left(\hat{G}^A_{00}\right)^{-1}\bullet\hat{G}^A_{0x^\prime}+\hat{G}^R_{x0}\bullet\left(\hat{G}^R_{00}\right)^{-1}\bullet\hat{G}^{<}_{00}\bullet\left(\hat{G}^A_{00}\right)^{-1}\bullet\hat{G}^A_{0x^\prime}.
\ee
For what concern the $R/A$ components of the Dyson's equation we have:
\be
\label{Boundary_RA_expression_met}
\hat{\overline{G}}^{R/A}_{xx^\prime}=\hat{G}^{R/A}_{xx^\prime}-\hat{G}^{R/A}_{x0}\bullet\left(\hat{G}^{R/A}_{00}\right)^{-1}\bullet\hat{G}^{R/A}_{0x^\prime}.
\ee
Finally, by using Eqs. (\ref{local_G}-\ref{nn_G}) and (\ref{x_x_prime_Lesser}) we obtain: 
\be
\label{Boundary_RA_expression_met_ex}
\hat{\overline{G}}(z)_{11\alpha\alpha}=\frac{z}{2t^2}\left(1-\sqrt{1-\left(\frac{2t}{z}\right)^2}\right)\mathbf{1},
\ee
where $R$ and $A$ are obtained by $z\to\omega\pm i0^+$, and 
\bal
\label{BC_xx_Lesser_Greater}
\hat{\overline{G}}^<_{11\alpha\alpha}(\omega)&=2i\pi\overline{\rho}(\omega)\left(\begin{array}{cc}f(\omega-\mu_\alpha) & 0 \\0 & f(\omega+\mu_\alpha) \end{array}\right),\\
\hat{\overline{G}}^>_{11\alpha\alpha}(\omega)&=-2i\pi\overline{\rho}(\omega)\left(\begin{array}{cc}1-f(\omega-\mu_\alpha) & 0 \\0 & 1-f(\omega+\mu_\alpha) \end{array}\right),
\eal
 with 
 \be
 \label{Semicircular_DOS}
 \overline{\rho}(\omega)=\theta(2t-|\omega|)\frac{\sqrt{4t^2-\omega^2}}{2\pi t^2}.
 \ee

\paragraph{Kitaev chain.}
Analogously to the previous case the starting point is a Kitaev chain with PBC, that in the $k$-space is described by the Hamiltonian:
\be
\mathcal{H}_{k}=\xi_{k}\sigma^{z}+\Delta_{k}\sigma^{y}
\ee
and $\xi_{k}=-2t\cos k-\mu$, $\Delta_{k}=2\Delta\sin k$. 
As a function of the complex variable $z$ the Green's function reads:
\be
\label{k_space_SC}
\hat{G}_{k}(z)=\sum_{\mu} \sigma^{\mu}G_{\mu k}(z),
\ee
where
\be
\label{k_space_SC_comp}
G_{0k}(z)=\frac{z}{(z)^2-(\xi^2_k+\Delta^2_k)},\,\,G_{yk}(z)=\frac{\Delta_k}{(z)^2-(\xi^2_k+\Delta^2_k)},\,\,G_{zk}(z)=\frac{\xi_k}{(z)^2-(\xi^2_k+\Delta^2_k)}.
\ee
By performing the Fourier transform Eq. (\ref{metallic_real_space}) of $\hat{G}_{k}(z)$ we obtain
\be
\label{SC_local_GR}
\hat{G}_{xx}(z)=\omega\sigma^0\mathcal{F}_{-1}(z)+\left(2t\mathcal{F}_{0}(z)-\mu\mathcal{F}_{-1}(z)\right)\sigma^z,
\ee
and 
\bal
\label{SC_nn_GR}
\hat{G}_{x+1x}(z)=\sum_{\mu}\sigma^{\mu}G_{\mu x+1x}(z),\quad\hat{G}_{xx+1}(z)=\sum_{\mu}\sigma^{\mu}G_{\mu xx+1}(z),
\eal 
where 
\bal
&G_{0x+1x}(z)=G_{0xx+1}(z)=-z\mathcal{F}_{0}(z),\\
&G_{2x+1x}(z)=-G_{2xx+1}(z)=2i\Delta\left(\frac{1}{4t^2-4\Delta^2}-\mathcal{F}_{1}(z)+\mathcal{F}_{-1}(z)\right),\\
&G_{3x+1x}(z)=G_{3xx+1}(z)=2t\left(\frac{1}{4t^2-4\Delta^2}-\mathcal{F}_{1}(z)\right)+\mu\mathcal{F}_{0}(z).
\eal
 In the previous expression we have introduced the quantity:
\be
\mathcal{F}_{m}(z)=\frac{1}{4(t^2-\Delta^2)}\frac{1}{Q_{+}(z)-Q_{-}(z)}\sum_{s=\pm1}\frac{sQ^{m}_s(z)}{\sqrt{1-1/Q^2_{s}(z)}},
\ee
and
\be
Q_{\pm}(\omega)=\frac{1}{2(\Delta^2-t^2)}\left[-t\mu\pm\sqrt{\Delta^2\mu^2-(\Delta^2-t^2)(z^2-4\Delta^2)}\right].
\ee
Analogously to the metallic case we introduce the perturbation (\ref{perturbation}), and by performing the perturbative expansion we obtain the Dyson's equation (\ref{Boundary_RA_expression_met}) for the $R$ and $A$ components of the boundary Green's functions. By using the local Green's function (\ref{SC_local_GR}) and the one evaluated at nearest-neighbor sites (\ref{SC_nn_GR}) we can easily compute $\hat{\overline{G}}^{R/A}_{11}(\omega)$.

\section{Bogoliubov-De Gennes equations}
\label{BdG}

\begin{figure}
\begin{centering}
\includegraphics[scale=0.5]{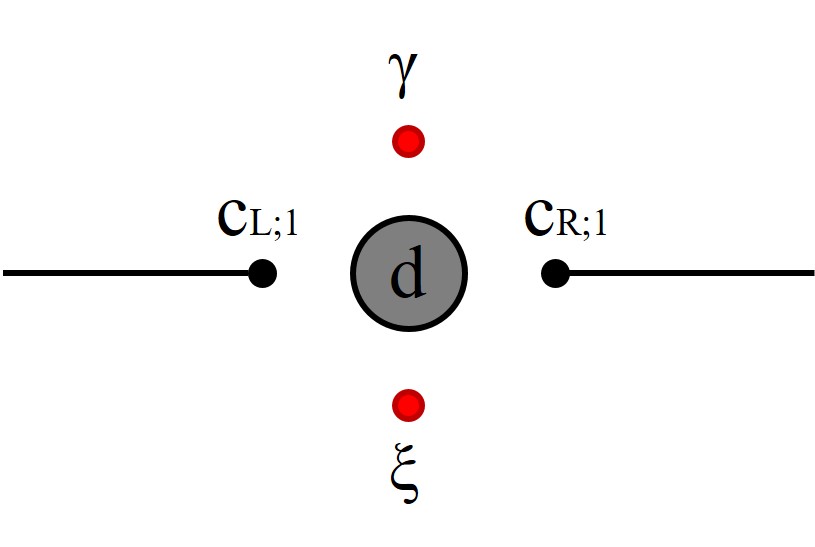}
\par\end{centering}
\caption{Sketch of the quantum dot ($d_{\sigma}$), connected to two fully-polarized
leads ($c_{L}$, $c_{R}$) and two Majorana fermion ($\eta$, $\xi$) }
\end{figure}

In this section we are interested to characterize the in-gap low-energy scattering processes \cite{Nazarov_Blanter}. Therefore, we approximate the Kitaev chains keeping only the MZM, $\gamma$ and $\xi$, so that the Hamiltonian (\ref{model_junction_SS_B}) takes the form: 
\bal
\label{scattering_H}
H=&-\sum_{j=1}^{\ell}\sum_{\alpha=L,R}t_{\alpha}\left(c_{\alpha j}^{\dagger}c_{\alpha j+1}+c_{\alpha j+1}^{\dagger}c_{\alpha j}\right)-\sum_{j=1}^{\ell}\sum_{\alpha=L.R}\mu_{\alpha}c_{\alpha j}^{\dagger}c_{\alpha j}\nonumber\\
 &-\sum_{\alpha=L,R}V_{\alpha}\left(c_{\alpha1}^{\dagger}d_{\uparrow}+d_{\uparrow}^{\dagger}c_{\alpha1}\right)-iV\gamma\gamma^d_{\uparrow}-i\tilde{V}\xi\xi^d_{\uparrow} \nonumber \\
 &+\frac{U}{4}\left(2d_{\downarrow}^{\dagger}d_{\downarrow}-1\right)\left(2d_{\uparrow}^{\dagger}d_{\uparrow}-1\right)\nonumber \\
 =&   H^{*}-\tilde{V}\xi\left(d_{\uparrow}-d_{\uparrow}^{\dagger}\right),
\eal
where, in comparison with Hamiltonian $H^*$ in Eq.(\ref{model_junction_SS_B}), we introduce an additional Majorana mode $\xi$ coupled to the impurity with hybridization $\tilde{V}$.

The Hamiltonian can be exactly solved looking for the solutions of the secular equation: 
\be
\left[\Gamma_{E,\delta},H\right]=E\Gamma_{E,\delta}.
\ee
Through the ansatz
\be
\Gamma_{E,\delta}=\sum_{j=1}^{\ell}\sum_{\alpha=L,R}\left(u_{\alpha;j}c_{\alpha j}+v_{\alpha;j}c_{\alpha j}^{\dagger}\right)+\left(u_{d}d_{\uparrow}+v_{d}d_{\uparrow}^{\dagger}\right)+\phi\gamma+\psi\xi,
\ee
the Bogoliubov-de Gennes (BdG) equations take the form
\begin{eqnarray}
-t_{\alpha}\left(u_{\alpha;j-1}+u_{\alpha;j+1}\right)-\mu_{\alpha}u_{\alpha;j} & = & Eu_{\alpha;j}\nonumber \\
t_{\alpha}\left(v_{\alpha;j-1}+v_{\alpha;j+1}\right)+\mu_{\alpha}v_{\alpha;j} & = & Ev_{\alpha;j}\label{eq:BdG-bulk}
\end{eqnarray}
within the bulk of the metallic leads, that is for $j>1$. At the
boundary BdG equations are 
\begin{eqnarray}
-t_{\alpha}u_{\alpha;2}-\mu_{\alpha}u_{\alpha;1}-V_{\alpha}u_{d} & = & Eu_{\alpha;1}\nonumber \\
t_{\alpha}v_{\alpha;2}+\mu_{\alpha}v_{\alpha;1}+V_{\alpha}v_{d} & = & Ev_{\alpha;1}\label{eq:BdG-c1}
\end{eqnarray}
for the endpoints of the leads,
\begin{eqnarray}
-V_{L}u_{L;1}-V_{R}u_{R;1}-2iV\phi-2\tilde{V}\psi+qUu_{d} & = & Eu_{d}\nonumber \\
V_{L}v_{L;1}+V_{R}v_{R;1}-2iV\phi+2\tilde{V}\psi-qUv_{d} & = & Ev_{d}\label{eq:BdG-d}
\end{eqnarray}
for the dot, and 
\begin{eqnarray}
iVv_{d}+iVu_{d} & = & E\phi\nonumber \\
\tilde{V}v_{d}-\tilde{V}u_{d} & = & E\psi\label{eq:BdG-Majorana}
\end{eqnarray}
for the two Majorana fermions. The solutions of the BdG equation inside
the bulk take the form 
\begin{equation}
\left(\begin{array}{c}
u_{\alpha;j}\\
v_{\alpha;j}
\end{array}\right)=\left(\begin{array}{c}
u_{\alpha}\\
v_{\alpha}
\end{array}\right)e^{ik_{\alpha}j}
\end{equation}
that inserted into Eq.(\ref{eq:BdG-bulk}) gives the secular equation
\begin{equation}
\left(\begin{array}{cc}
E+2t\cos\left(k_{\alpha}\right)+\mu_{\alpha} & 0\\
0 & E-2t\cos\left(k_{\alpha}\right)-\mu_{\alpha}
\end{array}\right)\left(\begin{array}{c}
u_{\alpha}\\
v_{\alpha}
\end{array}\right)=\left(\begin{array}{c}
0\\
0
\end{array}\right),
\end{equation}
and the dispersion relation 
\be
E^{2}-\left(2t\cos\left(k_{\alpha}\right)+\mu_{\alpha}\right)^{2}=0.
\ee
The latter equation admits four kind of waves (incoming particle in-p, outgoing
particle out-p, incoming hole in-h, outgoing hole out-h), such that
the most general eigenfunction with energy $E$ is given by

\begin{equation}
\left(\begin{array}{c}
u_{\alpha;j}\\
v_{\alpha;j}
\end{array}\right)=A_{p-in}^{\alpha}\left(\begin{array}{c}
1\\
0
\end{array}\right)e^{-ik_{p;\alpha}j}+A_{p-out}^{\alpha}\left(\begin{array}{c}
1\\
0
\end{array}\right)e^{ik_{p;\alpha}j}+A_{h-in}^{\alpha}\left(\begin{array}{c}
0\\
1
\end{array}\right)e^{ik_{h;\alpha}j}+A_{h-out}^{\alpha}\left(\begin{array}{c}
0\\
1
\end{array}\right)e^{-ik_{h;\alpha}j}
\end{equation}
with

\begin{eqnarray}
\cos k_{p;\alpha} & = & -\frac{E+\mu_{\alpha}}{2t}\nonumber \\
\cos k_{h;\alpha} & = & \frac{E-\mu_{\alpha}}{2t}.
\end{eqnarray}
The actual energy eigenstates are determined imposing the boundary
BdG equation. Scattering through the quantum dot junction is fully
encoded in the single-particle scattering matrix, $S$, that relates
the outgoing waves, $\bar{A}_{out}=\left(A_{p-out}^{1},A_{p-out}^{2},A_{h-out}^{1},A_{h-out}^{2},\right)^{t}$,
to the incoming waves, $\bar{A}_{in}=\left(A_{p-in}^{1},A_{p-in}^{2},A_{h-in}^{1},A_{h-in}^{2},\right)^{t}$.
Combining Eq.(\ref{eq:BdG-Majorana}) and Eq.(\ref{eq:BdG-d})
together we can write

\begin{equation}
\left(\begin{array}{c}
u_{d}\\
v_{d}
\end{array}\right)=\left(\begin{array}{cc}
Q_{+} & Q_{0}\\
Q_{0} & Q_{-}
\end{array}\right)\left(\begin{array}{c}
-V_{L}u_{L;1}-V_{R}u_{R;1}\\
V_{L}v_{L;1}V_{R}v_{R;1}
\end{array}\right)\label{eq:dot+majorana}
\end{equation}
with 
\begin{eqnarray}
Q_{\pm} & = & \frac{E\left(E^{2}\pm EqU+2V^{2}-2\tilde{V}^{2}\right)}{E^{4}-E^{2}q^{2}U^{2}+4E^{2}V^{2}-4E^{2}\tilde{V}^{2}-16V^{2}\tilde{V}^{2}},\nonumber \\
Q_{0} & = & -\frac{E\left(2V_{K:\eta}^{2}+2V_{K:\xi}^{2}\right)}{E^{4}-E^{2}q^{2}U^{2}+4E^{2}V^{2}-4E^{2}\tilde{V}^{2}-16V^{2}\tilde{V}^{2}}.
\end{eqnarray}
Inserting Eq.(\ref{eq:dot+majorana}) into Eq.(\ref{eq:BdG-c1}),
we finally arrive to 
\be
\left(\hat{M}\hat{\Phi}_{1}+\hat{t}\hat{\Phi}_{2}\right)\bar{A}_{out}+\left(\hat{M}\hat{\Phi}_{1}^{\dagger}+\hat{t}\hat{\Phi}_{2}^{\dagger}\right)\bar{A}_{in}=0,
\ee
where we have defined

\begin{equation}
\hat{M}=E\mathbb{I}+\sigma_{z}\otimes\left(\begin{array}{cc}
\mu_{1} & 0\\
0 & \mu_{2}
\end{array}\right)-\left(\begin{array}{cc}
Q_{+} & Q_{0}\\
Q_{0} & Q_{-}
\end{array}\right)\otimes\left(\begin{array}{cc}
V_{L}^{2} & V_{L}V_{R}\\
V_{L}V_{R} & V_{R}^{2}
\end{array}\right),
\end{equation}
and

\begin{equation}
\hat{t}=\sigma_{z}\otimes\left(\begin{array}{cc}
t_{1} & 0\\
0 & t_{2}
\end{array}\right),
\end{equation}
and 
\begin{equation}
\hat{\Phi}_{j}=\left(\begin{array}{cccc}
e^{ik_{p;\alpha}j} & 0 & 0 & 0\\
0 & e^{-ik_{p;\alpha}j} & 0 & 0\\
0 & 0 & e^{-ik_{h;\alpha}j} & 0\\
0 & 0 & 0 & e^{ik_{h;\alpha}j}
\end{array}\right).
\end{equation}
We have then 
\begin{eqnarray}
\bar{A}_{out} & = & \left[\left(\hat{M}\hat{\Phi}_{1}+\hat{t}\hat{\Phi}_{2}\right)^{-1}\left(\hat{M}\hat{\Phi}_{1}^{\dagger}+\hat{t}\hat{\Phi}_{2}^{\dagger}\right)\right]\bar{A}_{in}\nonumber \\
 & = & \hat{S}\left(E\right)\bar{A}_{in}.
\end{eqnarray}
The scattering matrix is an unitary matrix that encodes all the possible
single-particle processes at the junction with $E$ the energy of
the incoming particle/hole from the leads. Consistently with the notation
above, we have 
\begin{equation}
\hat{S}\left(E\right)=\left(\begin{array}{cccc}
r_{1,1}^{p,p} & t_{1,2}^{p,p} & a_{1,1}^{p,h} & c_{1,2}^{p,h}\\
t_{2,1}^{p,p} & r_{2,2}^{p,p} & c_{2,1}^{p,h} & a_{2,2}^{p,h}\\
a_{1,1}^{h,p} & c_{1,2}^{h,p} & r_{1,1}^{h,h} & t_{1,2}^{h,h}\\
c_{2,1}^{h,p} & a_{2,2}^{h,p} & t_{2,1}^{h,h} & r_{2,2}^{h,h}
\end{array}\right),
\end{equation}
where $r_{\alpha,\alpha}^{\mu,\mu}\left(E\right)$ denotes the reflection
amplitude of a particle or of an hole, $t_{\alpha,\bar{\alpha}}^{\mu,\mu}\left(E\right)$
is the trasmission amplitude between the leads, $a_{\alpha,\alpha}^{\mu,\bar{\mu}}\left(E\right)$
corresponds to the Andreev reflection, that is the conversion of a
particle (hole) into an hole (particle) within the same lead, finally
$c_{\alpha,\bar{\alpha}}^{\mu,\bar{\mu}}\left(E\right)$ is the crossed
Andreev reflection amplitude, that is the conversion of a particle
(hole) in one lead to an hole (particle) in the other lead. The scattering
matrix allows us to introduce the four kind of eigenstates that define
the scattering states basis. We have:

(i) incoming particle from left lead (pL) 
\begin{eqnarray}
\left(\begin{array}{c}
u_{L;j}^{pL}\\
v_{L;j}^{pL}
\end{array}\right) & = & A_{pL}\left(\begin{array}{c}
e^{-ik_{p,1}j}+r_{1,1}^{e,e}e^{ik_{p,1}j}\\
a_{1,1}^{h,e}e^{-ik_{h,1}j}
\end{array}\right)\nonumber \\
\left(\begin{array}{c}
u_{R;j}^{pL}\\
v_{R;j}^{pL}
\end{array}\right) & = & A_{pL}\left(\begin{array}{c}
t_{2,1}^{e,e}e^{ik_{p,2}j}\\
c_{2,1}^{h,e}e^{-ik_{h,2}j}
\end{array}\right);
\end{eqnarray}

(ii) incoming particle from right lead (pR)
\begin{eqnarray}
\left(\begin{array}{c}
u_{L;j}^{pR}\\
v_{L;j}^{pR}
\end{array}\right) & = & A_{pR}\left(\begin{array}{c}
t_{1,2}^{e,e}e^{ik_{p,1}j}\\
c_{1,2}^{h,e}e^{-ik_{h,1}j}
\end{array}\right)\nonumber \\
\left(\begin{array}{c}
u_{R;j}^{pR}\\
v_{R;j}^{pR}
\end{array}\right) & = & A_{pR}\left(\begin{array}{c}
e^{-ik_{p,2}j}+r_{2,2}^{e,e}e^{ik_{p,2}j}\\
a_{2,2}^{h,e}e^{-ik_{h,2}j}
\end{array}\right);
\end{eqnarray}

(iii) incoming hole from left lead (hL) 
\begin{eqnarray}
\left(\begin{array}{c}
u_{L;j}^{hL}\\
v_{L;j}^{hL}
\end{array}\right) & = & A_{hL}\left(\begin{array}{c}
a_{1,1}^{e,h}e^{ik_{p,1}j}\\
e^{ik_{h,1}j}+r_{1,1}^{h,h}e^{-ik_{h,1}j}
\end{array}\right)\nonumber \\
\left(\begin{array}{c}
u_{R;j}^{pL}\\
v_{R;j}^{pL}
\end{array}\right) & = & A_{hL}\left(\begin{array}{c}
c_{2,1}^{e,h}e^{ik_{p,2}j}\\
t_{2,1}^{h,h}e^{-ik_{h,2}j}
\end{array}\right);
\end{eqnarray}

(iv) incoming hole from right lead (hR)
\begin{eqnarray}
\left(\begin{array}{c}
u_{L;j}^{hR}\\
v_{L;j}^{pR}
\end{array}\right) & = & A_{hR}\left(\begin{array}{c}
c_{1,2}^{e,h}e^{ik_{p,1}j}\\
t_{1,2}^{h,h}e^{-ik_{h,1}j}
\end{array}\right)\nonumber \\
\left(\begin{array}{c}
u_{R;j}^{hR}\\
v_{R;j}^{pR}
\end{array}\right) & = & A_{hR}\left(\begin{array}{c}
a_{2,2}^{e,h}e^{ik_{p,2}j}\\
e^{ik_{h,2}j}+r_{2,2}^{h,h}e^{-ik_{h,2}j}
\end{array}\right);
\end{eqnarray}
with $A_{\delta}$ appropriate normalization constants. In the following,
to simplify the notation, we will assume particle-hole symmetry, $S_{\alpha,\beta}^{\mu,\lambda}\left(E\right)=\left[S_{\alpha,\beta}^{\lambda,\mu}\left(-E\right)\right]^{*}$,
and assume the junction to be symmetric respect the lead exchange,
$S_{\alpha,\beta}^{\mu,\lambda}\left(E\right)=S_{\beta,\alpha}^{\mu,\lambda}\left(E\right)$.
Because of these symmetries, we have only four relevant scattering
coefficients, $\left|S_{i,j}\right|^{2}$, that fully describe the
physics at the junction. We refer to them as $R\left(E\right)$, normal
reflection, $T\left(E\right)$, normal transmission, $A\left(E\right)$,
Andreev reflection and $C\left(E\right)$, crossed Andreev reflection.
It is important to highlight that the normal transmission and the
Andreev reflection are the only processes that creates an imbalance
in the relative number of particles within the two metallic contacts. Whereas Andreev
reflection and crossed Andreev reflection do not preserve the total
number of particle in the metallic lead subsystem, as shown in Eq.(\ref{charge_imbalance})

\begin{eqnarray}
\label{charge_imbalance}
R & \rightarrow & \frac{\left(\dot{N}_{L}-\dot{N}_{R}\right)}{2}=0;\ \ \ \frac{\left(\dot{N}_{L}+\dot{N}_{R}\right)}{2}=0;\nonumber \\
T & \rightarrow & \frac{\left(\dot{N}_{L}-\dot{N}_{R}\right)}{2}=1;\ \ \ \frac{\left(\dot{N}_{L}+\dot{N}_{R}\right)}{2}=0;\nonumber \\
A & \rightarrow & \frac{\left(\dot{N}_{L}-\dot{N}_{R}\right)}{2}=1;\ \ \ \frac{\left(\dot{N}_{L}+\dot{N}_{R}\right)}{2}=-1;\nonumber \\
C & \rightarrow & \frac{\left(\dot{N}_{L}-\dot{N}_{R}\right)}{2}=0;\ \ \ \frac{\left(\dot{N}_{L}+\dot{N}_{R}\right)}{2}=-1;
\end{eqnarray}
In Fig.(2) we report the scattering coefficients in presence
of zero, one and two Majorana fermions. 
\begin{figure}
\begin{centering}
\includegraphics[scale=0.5]{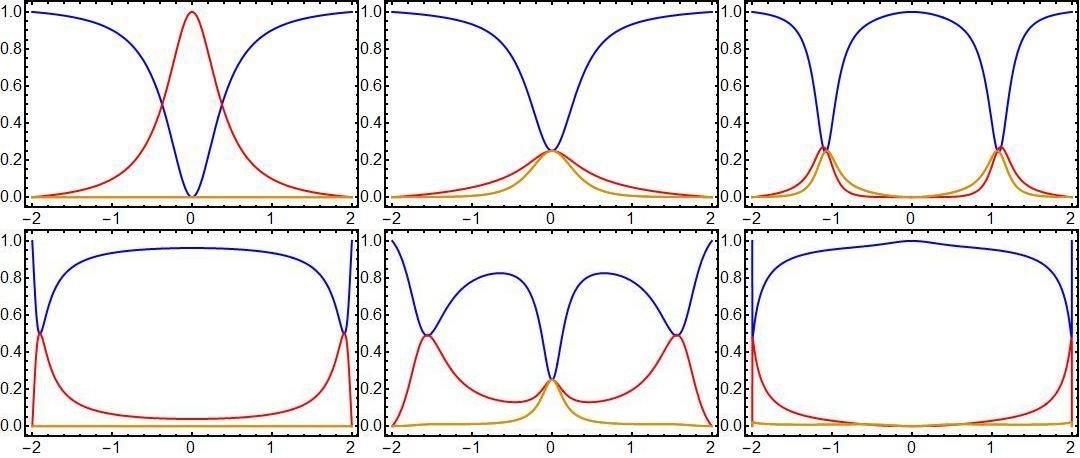}
\par\end{centering}

\caption{Scattering matrix coefficients for a quantum dot coupled with a) zero,
b) one or c) two Majorana fermions and for zero (first row) or finite,
$U=1.6$, interaction (second row). Blue: normal reflection; red:
normal transmission, orange: Andreev reflection and crossed Andreev
reflection.}
\end{figure}
 In the absence of Majorana fermions and for $U=0$,
the junction is trasparent, in the zero energy limit, due the resonance
with the zero energy quantum dot states. However, for any finite $U$,
the resonance is suppressed by the interaction that removes low energy states in the quantum dot. 
On the other hand, in the presence of
a MZM we observe a zero bias trasmission and Andreev peaks that persist even for large values of $U$. 
Finally, in the presence of two Majorana, no
zero energy state survives due to the hybridization between the Kitaev chains and
the quantum dot. The robust topological peak in the scattering matrix
coefficients is then expected to be an interesting signature of the
presence of a MZM. In the following we will relate
these features to physically measurable quantities like the current
and the shot noise.

To spell out the relation between the scattering matrix amplitudes
and the current, it us useful to
express the fermionic creation and annihilation operators in real
space as a function of the system eigenvectors

\begin{eqnarray}
\label{c-Gamma}
c_{\alpha j} & = & \sum_{E>0}\sum_{\delta}\left(\left[u_{\alpha;j}^{\delta,E}\right]^{*}\Gamma_{\delta,E}+\left[v_{\alpha;j}^{\delta,E}\right]^{*}\Gamma_{\delta,E}^{\dagger}\right)\nonumber \\
c_{\alpha j}^{\dagger} & = & \sum_{E>0}\sum_{\delta}\left(\left[u_{\alpha;j}^{\delta,E}\right]\Gamma_{\delta,E}^{\dagger}+\left[v_{\alpha;j}^{\delta,E}\right]\Gamma_{\delta,E}\right),
\end{eqnarray}
with $\delta$ running over the four scattering states. The eigenvectors
satisfy the fermionic algebra

\begin{equation}
\left\{ \Gamma_{\delta,E},\Gamma_{\delta',E'}^{\dagger}\right\} =\delta_{\delta,\delta'}\delta_{E,E'}
\end{equation}
all other anticommutator vanish.

The Landauer-Buttiker approach, that consist in shooting particles
and holes agains the junction from thermal reservoirs at fixed temperature
and voltage biased chemical potentials, allows us to express the transport
properties of the systems in terms of the voltage bias into the leads
and the scattering matrix amplitudes. The starting point is the current
operator in lead $\alpha$, defined as

\begin{equation}
J_{\alpha j}=-iet\left(c_{\alpha j}^{\dagger}c_{\alpha j+1}-c_{\alpha j+1}^{\dagger}c_{\alpha j}\right).
\end{equation}
Expressing the fermionic operator in terms of the system eigenvectors
we obtain

\begin{eqnarray}
\left\langle J_{\alpha j}\left(t\right)\right\rangle & = &  -iet\sum_{E,E'>0}\sum_{\delta,\delta'}\left\{ \left\langle \Gamma_{\delta,E}^{\dagger}\Gamma_{\delta',E'}\right\rangle e^{i\left(E-E'\right)t}\left[u_{\alpha;j}^{\delta,E}\left(u_{\alpha;j+1}^{\delta',E'}\right)^{*}-u_{\alpha;j+1}^{\delta,E}\left(u_{\alpha;j}^{\delta',E'}\right)^{*}\right]\right.\nonumber \\
 &  & \left\langle \Gamma_{\delta,E}\Gamma_{\delta',E'}^{\dagger}\right\rangle e^{-i\left(E-E'\right)t}\left[v_{\alpha;j}^{\delta,E}\left(v_{\alpha;j+1}^{\delta',E'}\right)^{*}-v_{\alpha;j+1}^{\delta,E}\left(v_{\alpha;j}^{\delta',E'}\right)^{*}\right]\nonumber \\
 &  & \left\langle \Gamma_{\delta,E}\Gamma_{\delta',E'}\right\rangle e^{i\left(E+E'\right)t}\left[v_{\alpha;j}^{\delta,E}\left(u_{\alpha;j+1}^{\delta',E'}\right)^{*}-v_{\alpha;j+1}^{\delta,E}\left(u_{\alpha;j}^{\delta',E'}\right)^{*}\right]\nonumber \\
 &  & \left.\left\langle \Gamma_{\delta,E}^{\dagger}\Gamma_{\delta',E'}^{\dagger}\right\rangle e^{-i\left(E+E'\right)t}\left[u_{\alpha;j}^{\delta,E}\left(v_{\alpha;j+1}^{\delta',E'}\right)^{*}-u_{\alpha;j+1}^{\delta,E}\left(v_{\alpha;j}^{\delta',E'}\right)^{*}\right]\right\},
\end{eqnarray}
that does not depend on the site index $j$.
In the presence of a chemical potential bias, particles emerges from left reservoir at 
$\mu_{L}=-\phi/2$ and at $\mu_{R}=\phi/2$ on the
right lead:

\begin{eqnarray}
\left\langle \Gamma_{pL,E}^{\dagger}\Gamma_{pL,E}\right\rangle  & = & f\left(E-\frac{\phi}{2}\right)\nonumber \\
\left\langle \Gamma_{pR,E}^{\dagger}\Gamma_{pR,E}\right\rangle  & = & f\left(E+\frac{\phi}{2}\right)\nonumber \\
\left\langle \Gamma_{hL,E}^{\dagger}\Gamma_{hL,E}\right\rangle  & = & f\left(E+\frac{\phi}{2}\right)\nonumber \\
\left\langle \Gamma_{hR,E}^{\dagger}\Gamma_{hR,E}\right\rangle  & = & f\left(E-\frac{\phi}{2}\right)
\end{eqnarray}
where $f\left(E\right)$ is the Fermi distribution function and all
the other expectation values vanish. We finally obtain

\begin{eqnarray}
\left\langle\left\langle J_{Q} \right\rangle\right\rangle=\frac{\left\langle\left\langle J_{L} \right\rangle\right\rangle-\left\langle\left\langle J_{R} \right\rangle\right\rangle}{2} & = & \frac{e^2}{h}\int dE\left[f\left(E-\frac{\phi}{2}\right)-f\left(E+\frac{\phi}{2}\right)\right]\left[A\left(E\right)+T\left(E\right)\right]\nonumber \\
 & = & \frac{e^2}{h}\int dE\left[f\left(E-\frac{\phi}{2}\right)-f\left(E+\frac{\phi}{2}\right)\right]\left[1-R\left(E\right)+C\left(E\right)\right].
\end{eqnarray}
As a function of the number of Majorana connected to the leads, referring
to Fig.(2), we have, for an interacting dot
\begin{eqnarray}
N_{MZM}=0 & \rightarrow & \left\langle\left\langle J_{Q} \right\rangle\right\rangle=\frac{e^2}{h} \phi\left(\frac{4\pi \bar{\rho}(0) V^2_c}{U}\right)^2/\left[1+\left(\frac{4\pi \bar{\rho}(0) V^2_c}{U}\right)^2\right] \nonumber \\
N_{MZM}=1 & \rightarrow & \left\langle\left\langle J_{Q} \right\rangle\right\rangle=\frac{e^2}{2h}\phi\\
N_{MZM}=2 & \rightarrow & \left\langle\left\langle J_{Q} \right\rangle\right\rangle=0\nonumber,
\end{eqnarray}
in agreement with the Keldysh results. By inserting Eq.(\ref{c-Gamma}) in Eq.(\ref{noise}) we obtain the expression of the shot noise in terms of the scattering matrix amplitudes. The final result coincides with Eq.(\ref{zerotemperature_shotnoise}).

\bibliographystyle{apsrev}
\bibliography{mybiblio}